\documentclass[%
 reprint,
 amsmath,amssymb,
 aps,
]{revtex4-2}
\usepackage{amssymb}
\usepackage{amsmath}
\usepackage{hyperref}
\usepackage{graphicx}
\usepackage{parskip}
\usepackage{caption}
\usepackage{lipsum}
\usepackage{dcolumn}
\usepackage{bm}
\usepackage{float}
\usepackage{subcaption}
\usepackage{upgreek}
\usepackage{xurl}
\usepackage{tabularx}
\usepackage{makecell}
\allowdisplaybreaks
\begin{document}
\preprint{APS/123-QED}
\title{A mean-field theory approach to 3D nematic phase transitions in microtubules}
\author{Cameron Gibson$^{\text{1,2}}$}
\email{camerongibson@tamu.edu}
\author{Henrik J$\text{\"{o}}$nsson$^{\text{1,3,4}}$}
\email{Co-corresponding author: henrik.jonsson@slcu.cam.ac.uk}
\author{Tamsin A. Spelman$^{\text{1}}$}
\email{Co-corresponding author: tamsin.spelman@slcu.cam.ac.uk}

\affiliation{
$^\text{1}$ Sainsbury Laboratory, University of Cambridge, Cambridge, UK.\\
$^\text{2}$ Department of Physics and Astronomy, Texas A\&M University, College Station, Texas 77843. (Current Addr.)\\
$^\text{3}$ Department of Applied Mathematics and Theoretical Physics, University of Cambridge, Cambridge, UK.\\
$^\text{4}$ Center for Environmental and Climate Research, Lund University, Lund, Sweden.}
\date{\today}
\begin{abstract}
	    Microtubules are dynamic intracellular fibers that have been observed experimentally to undergo spontaneous self-alignment. We formulate a 3D mean-field theory model to analyze the nematic phase transition of microtubules growing and interacting within a 3D space then make a comparison with computational simulations. We identify a control parameter $G_\text{eff}$ and predict a unique critical value $G_\text{eff}=1.56$ for which a phase transition can occur. Furthermore, we show both analytically and using simulations that this predicted critical value does not depend on the presence of zippering. The mean-field theory developed here provides an analytical estimate of microtubule patterning characteristics without running time-consuming simulations and is a step towards bridging scales from microtubule behavior to multicellular simulations.
\end{abstract}
\maketitle
	\pagenumbering{roman}
	\newpage
	\pagenumbering{arabic}

\section{Introduction}
Microtubules are long filamentous fibers found in all eukaryotic cells \cite{micro} and are vital for many processes at the cell level that are in turn essential for the survival and development of cells and the larger organism \cite{greenbook}. These processes include cell expansion and division \cite{division,expansion,morpho}; internal transportation such as nucleus repositioning before cell division or cellulose deposition to grow cells \cite{5,6,7}; fertilization \cite{8}; and providing mechanical structure in animal cells \cite{greenbook}.

Microtubules form one part of the cytoskeleton (the intracellular dynamic fiber network) that also consists of actin fibres and, in animal cells, intermediate filaments \cite{cyto,review}. Microtubules continuously grow and shrink via the assembly and disassembly of the protein tubulin \cite{dinst}. They undergo local stochastic behaviors such as spontaneous catastrophe, rescue, and nucleation. Microtubules interact with each other displaying behaviors via zippering predominantly at small angle interactions, induced catastrophe predominantly at large angle interactions \cite{angles,boundary}, and crossover severing \cite{severing}. These complex behaviors make them very interesting systems to study from both a physical and mathematical perspective.

Microtubules typically nucleate from $\gamma$-tubulin complexes found on centrosomes in animal systems \cite{4} or from the cortex in plant cells \cite{3}. However, microtubules have also been observed nucleating in the cytoplasm of neurons \cite{1} and the moss \emph{Physcomitrella patens} \cite{2}, as well as nucleating by branching off existing microtubules \cite{branching}, demonstrating the need for a microtubule model incorporating 3D microtubule nucleation and orientation.

Microtubule networks can be viewed as analogous to the condensed matter system of nematic liquid crystals \cite{liqc,liqcpedagog} as they can both be described as systems of many hard interacting rods. Furthermore, high levels of spontaneous alignment have been observed experimentally in microtubule systems which are qualitatively similar to phase transitions in nematic liquid crystals \cite{experiment}. A primary use of this comparison has been the standard use of the nematic order parameter as a measure of the orientational alignment of microtubules, which will be used in this paper.

Many different computational models have been used to simulate microtubule dynamics \cite{comp1,comp2,comp4}. CorticalSim \cite{corticalsimwebsite,comp3} is an example of an efficient event-driven model for modelling microtubules restricted to a plane. It has been used to show, for example, that the co-alignment of microtubules nucleating from parent microtubules supports whole network alignment \cite{14}. A different model Cytosim \cite{cytosimwebsite} is a 3D force-based microtubule model used, for example, to consider how molecular motor patterns can direct filament directions \cite{11}. A third example is Tubulaton \cite{tubulatonwebsite,tubulaton}, a 3D rule-based model used, for example, to study the importance of the crossover-severing protein katanin to microtubule ordering in plant protoplasts, as observed in experiments \cite{durand}.

Similarly, several mathematical models have been proposed to analyze cytoskeletal dynamics \cite{model1,model2,model3,model5}. One useful continuum theory approach is mean-field theory, which is used extensively to model condensed matter systems \cite{mamod}. Mean-field theory in the context of cytoskeletal dynamics was to our knowledge introduced by Dogterom and Leibler \cite{mft}. They derived governing differential equations which incorporated the fundamental microtubule properties of growing, shrinking, catastrophe and rescue. This model was later expanded to include more complex microtubule behaviors\cite{CM,mft,Simons,mft1,baulin,main}. This was extended to the first 2D mean-field theory model \cite{baulin}, with subsequent models introducing more complex microtubule behaviors such as induced catastrophes in Hawkins \textit{et. al.} \cite{main}, where they showed the existence of a phase transition under certain assumptions.

 There is only one extension of mean-field to 3D of which the authors are aware \cite{thesis}. That model is restricted to the specific case of microtubules only nucleating radially from a prescribed central centrosome within a bounded domain, with the model including interaction dynamics between microtubules and the cell boundary but not between microtubules themselves. 
 In this paper, the 2D mean-field model of \cite{main} is extended to 3D in a novel way incorporating microtubule interaction dynamics in the different setup of microtubules nucleating randomly within a 3D domain. The differences and similarities between 2D and 3D are then highlighted and the theoretical predictions of the 3D model are compared to results obtained from 3D simulations from Tubulaton \cite{tubulatonwebsite}. This paper is organized as follows. The models are outlined in Sec.~\ref{section:section_one}. Specifically in Sec.~\ref{subsection:MFT}, the 3D mean-field theory model is derived and in Sec.~\ref{subsection:simulation}, the computational model Tubulaton used to validate the mean-field model is described. The results are presented in Sec.~\ref{section:section_two}. In Sec.~\ref{subsection:twosix}, the constraints on the system that allow for a phase transition in 3D are determined. In Sec.~\ref{subsection:positivec}, we compare our 3D mean-field model with the previous 2D mean-field theory model from Hawkins \textit{et. al.} \cite{main}. Finally, the predictions of the mean-field theory model are compared to the results of the computational simulations in Sec.~\ref{subsection:compare}, with the effects of severing considered in Sec.~\ref{subsection:Severing}, and to experimental values from the literature in Sec.~\ref{subsection:ExperimentComparison} before concluding in Sec.~\ref{section:conclusion}.

\section{Models}
\label{section:section_one}
Here, the 3D mean-field mathematical model is formulated and the computational model is briefly outlined. Throughout this paper, spherical polar coordinates are used to describe directions in 3D space, with $(\theta,\phi)$ representing polar and azimuthal angles, respectively. 
\subsection{Mean-Field Theory} \label{subsection:MFT}
Each microtubule is modelled as a series of segments. It is assumed that unhindered microtubules grow in a straight line, but can change direction with a prescribed angle-dependent probability when two microtubules collide. When the microtubule changes direction, the old segment becomes static (neither growing nor shrinking) and a new segment starts growing in the new direction, anchored to the previous segment. The microtubules grow (and shrink) in segments, with joints allowing each segment to be oriented in a different direction.

It is assumed that that microtubules isotropically nucleate everywhere in 3D space at a constant rate $r_{\text{n}}$, initiating in a growing state with static minus end and growing plus end. Microtubules are always static at the minus end and are either growing or shrinking at the plus end with speed $v^\text{+}$ or $v^\text{--}$, respectively. The plus end changes from shrinking to growing via spontaneous rescue with rate $r_{\text{r}}$ and changes from growing to shrinking via spontaneous catastrophe with rate $r_{\text{c}}$.

\begin{figure}
    \vspace{0cm}
    \centering
    \captionsetup{skip=-7.5cm,margin={0cm,0cm},justification=raggedright}
    \includegraphics[width=.45\textwidth,height=14cm]{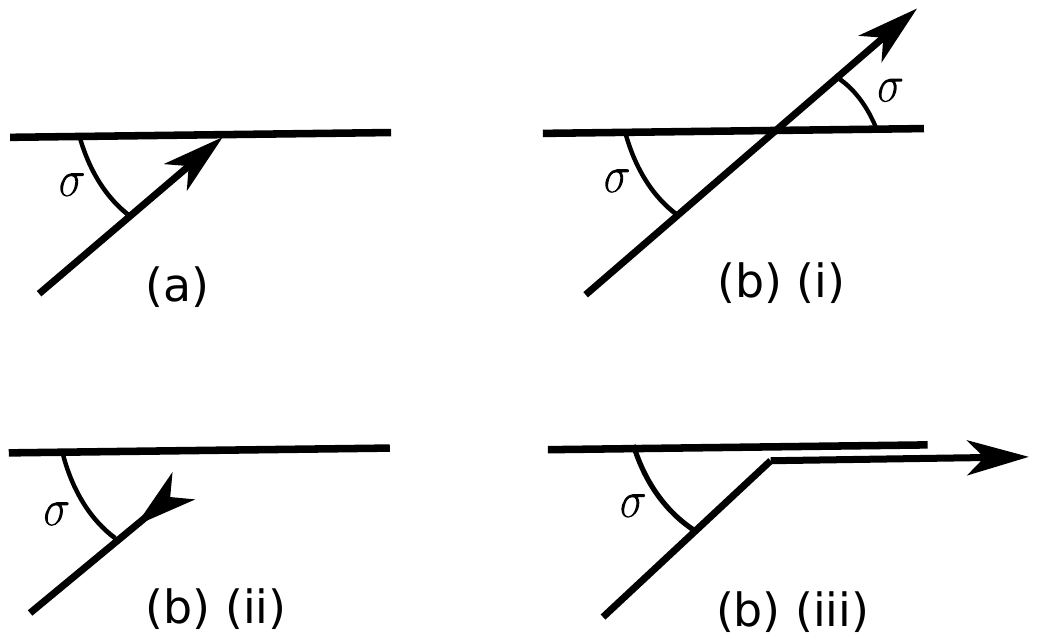}
    \caption{Illustration of different microtubule interactions. These behaviors are included in both the mean-field model and the simulations. (a) Initial collision where the growing microtubule segment collides with another microtubule at angle $\sigma$. (b) Different responses to the collision: (i) Crossover: the growing segment continues to grow unhindered. (ii) Induced catastrophe: the segment switches from growing to shrinking. (iii) Zippering: the segment starts to grow parallel to the segment with which it collides.}
    \label{drawing}
    \vspace{-.5cm}
\end{figure}

 When a growing segment collides with another microtubule, either there is an induced catastrophe (it starts shrinking), crossover (it keeps growing unhindered) or zippering (it starts growing parallel to the second segment) with respective probabilities $P_{\text{c}}(\sigma)$, $P_{\text{x}}(\sigma)$ and $P_{\text{z}}(\sigma)$ all written as functions of the collision angle $\sigma$ (Fig.~\ref{drawing}).
\subsubsection{Master Equations}
\label{subsection:me}
In this subsection, the governing 3D mean-field differential equations are derived, following a similar argument to that outlined in detail for a 2D framework \cite{main}.

The density of microtubules is assumed to be large enough for this discrete system to be accurately approximated by continuous variables. Therefore $m_{\text{i}}^{\text{+/--/0}}(l,\theta,\phi,t)$ is defined as the density of microtubule segments in direction $(\theta,\phi)$ of length $l$ at time $t$ with $\text{+/--/0}$ indicating a growing/shrinking/inactive segment, respectively, and $\text{i}$ indexing the segments (letting $\text{i}=1$ index the segment that has nucleated).

Then, the master equations governing the evolution of the system can be expressed in terms of flux terms denoted $\Phi_{\text{event}}$ as
\begin{equation}
\begin{split}
	\partial_t m_\text{i}^\text{+}(l,\theta,\phi,t) =& \Phi_{\text{growth}}+\Phi_{\text{rescue}}-\Phi_{\text{spontcat}}\\
	&-\Phi_{\text{inducedcat}}-\Phi_{\text{zipper}},\\
	\partial_t m_\text{i}^\text{--}(l,\theta,\phi,t) =& \Phi_{\text{shrinkage}}-\Phi_{\text{rescue}}+\Phi_{\text{spontcat}}\\
	&+\Phi_{\text{inducedcat}}+\Phi_{\text{reactivation}},\\
	\partial_t m_\text{i}^\text{0}(l,\theta,\phi,t) =& \Phi_{\text{zipper}}-\Phi_{\text{reactivation}},	\label{ss}
\end{split}
\end{equation}
where $\partial_x$ denotes partial differentiation with respect to $x$. Explicit expressions for the fluxes corresponding to behaviors independent of microtubule interactions follow from physical definitions as
\begin{equation}
\begin{split}
& \Phi_{\text{growth}}  \equiv \left(\partial_t l \right)  (\partial_l m_\text{i}^\text{+}(l,\theta,\phi,t))= -v^\text{+} \partial_l m_\text{i}^\text{+}(l,\theta,\phi,t), \\
   & \Phi_{\text{shrinkage}} \equiv \left(\partial_t l\right)  (\partial_l m_\text{i}^\text{--}(l,\theta,\phi,t)) = v^\text{--} \partial_l m_\text{i}^\text{--}(l,\theta,  \phi, t),\\
   & \Phi_{\text{rescue}} \equiv r_{\text{r}}m_\text{i}^\text{--}(l,\theta,\phi,t),\\
   & \Phi_{\text{spontcat}} \equiv r_{\text{c}}m_\text{i}^\text{+}(l,\theta,\phi,t).
\end{split}
\end{equation}
Flux terms associated with microtubule interactions have more complex formulations. The reactivation flux can be written as
\begin{equation}
\begin{split}
    &\Phi_{\text{reactivation}} \equiv v^\text{--} \int d \theta' \int d \phi' \sin(\theta') \\
    &\times m_\text{i+1}^\text{--} (l' =0,\theta',\phi', t) p_{\text{unzip}} (\theta,\phi,l| \theta', \phi' ,t),
    \end{split}
\end{equation}
where $p_{\text{unzip}}(\theta,\phi,l| \theta',\phi', t)$ is the probability that the $(i+1)^{\text{th}}$ segment shrinks from the direction $(\theta',\phi')$ to length $l'=0$ at time t and reactivates the $i^{\text{th}}$ segment of length $l$ in direction $(\theta,\phi)$. Detailed arguments as to why $p_{\text{unzip}}$ does not contribute in the steady state case can be found in \cite{main}, which naturally extend from 2D to 3D.

The length density of microtubules pointing in direction $(\theta,\phi)$ is 
\begin{equation}
\begin{split}
    k(\theta,\phi,t) \equiv \sum_{\text{i}=1}^\infty \int dl &\left[m_\text{i}^\text{+}(l,\theta,\phi, t)+m_\text{i}^\text{--}(l,\theta,\phi,t)\right.\\
    &\left.+m_\text{i}^\text{0}(l,\theta,\phi,t)\right]l.
    \end{split}
\end{equation}
The diameter of the microtubules is defined as $d_\text{m}$. The induced catastrophe flux term is given by
\begin{equation}
\begin{split}
\Phi_{\text{inducedcat}}\equiv & d_\text{m} v^\text{+} m_\text{i}^\text{+} (l,\theta,\phi,t) \int d\theta' \int d\phi' \sin(\theta')\\
&\times c(\theta,\theta',\phi-\phi') k(\theta',\phi',t) \label{IC},
\end{split}
\end{equation}
where it is defined that
\begin{equation}
     c(\theta,\theta',\phi-\phi') \equiv |\sin(\sigma)|P_\text{c}(\sigma(\theta,\theta',\phi-\phi')) \label{factor},
\end{equation}
where the angle between the two directions $(\theta,\phi)$ and $(\theta',\phi')$ has been denoted as
\begin{equation}
    \sigma \equiv \arccos(\sin(\theta)\sin(\theta')\cos(\phi-\phi')+\cos(\theta) \\ \cos(\theta')) \label{pw}.
\end{equation}

\begin{figure}
    \vspace{0cm}
    \hspace*{-.72cm}
    \captionsetup{skip=-.5cm,margin={0cm,0cm},justification=raggedright}
    \includegraphics[width=.55\textwidth,height=6cm]{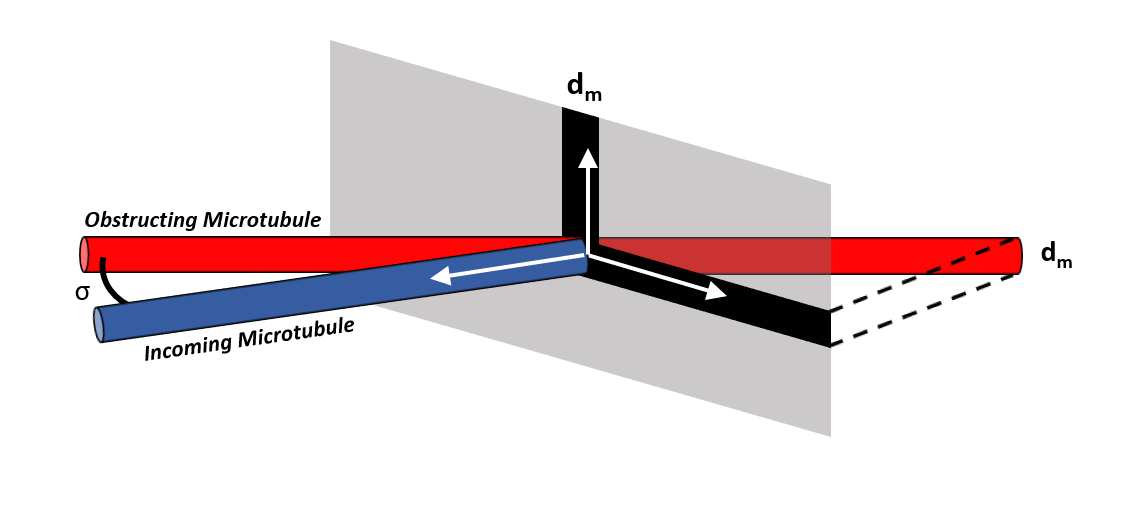}
    \caption{A graphical representation of Eq.~\ref{IC}. A second microtubule obstructs an incoming microtubule at an angle of $\sigma$. When the obstructing microtubule is projected onto the plane orthogonal to the incoming microtubule, we consider collisions as occurring in a length of the microtubule diameter $d_\text{m}$ in the direction orthogonal to both microtubules.}
    \label{projection}
    \vspace{0cm}
\end{figure}

This flux term represents the rate at which a microtubule of length $l$ growing in direction $(\theta,\phi)$ collides with an obstructing microtubule oriented in any direction. A significant difference from the 2D case is the factor of $d_\text{m}$ in Eq.~\ref{IC}. In 2D, two thin infinite lines will always collide if they are not parallel. However, in 3D, two thin non-parallel rods can pass over each other without colliding. This difference in the 3D case can be addressed by considering two microtubules to collide when they are within a distance of the microtubule diameter $d_\text{m}$ of each other in the direction orthogonal to both microtubules (Fig.~\ref{projection}). The $|\sin(\sigma)|$ factor in Eq.~\ref{factor} projects the length density $k(\theta',\phi')$ to a plane perpendicular to the direction $(\theta,\phi)$ of an incoming microtubule. The microtubule diameter $d_\text{m}$ is included as a distance in the direction perpendicular to both microtubules below which two microtubules interact. Note that the introduction of a factor of $d_\text{m}$ in the flux term ensures that both sides of Eq.~\ref{IC} have the same dimensionality ($\text{Length}^{-4} \times \text{Time}^{-1}$). 

The flux term $\Phi_{\text{zipper}}$ is defined similarly to $\Phi_{\text{inducedcat}}$ in Eq.~\ref{IC} but with $c$ replaced by $z$ (each of which contains a factor of $P_\text{c}$ and $P_\text{z}$, respectively). 
\subsubsection{Control Parameter}
\label{subsection:cp}
An important parameter of the system \cite{baulin,main} is
\begin{equation}g=r_{\text{r}}/v^\text{--} - r_{\text{c}}/v^\text{+}. \label{defcp}
\end{equation}
Physically, $g$ corresponds to the non-interacting behavior of the microtubules. The limit $g \rightarrow - \infty$ corresponds to the average length of the microtubules tending to zero resulting in a completely isotropic system, whilst $g \rightarrow \infty$ corresponds to the average length of the microtubules tending to infinity resulting in a completely ordered (anisotropic) system. Following the earlier analogy to a phase transition in a liquid crystal system, this control parameter $g$ is analogous to temperature in liquid crystals \cite{liqc,liqcpedagog}. In 2D, the existence of an orientational phase transition as $g$ increases from negative infinity has been shown \cite{main}. Importantly, both here and in 2D, the only physically realizable values of $g$ are negative, since positive $g$ corresponds to unbounded growth.
\subsubsection{Steady State System}
\label{subsection:sss}
Here, the previously derived master equations (Eq.~\ref{ss}) are reduced to the steady state case. The arguments are similar to the 2D derivation \cite{main}.

Firstly, for simplification it is noted expressions like Eq.~\ref{IC} can be summarized by defining the dimensionless linear operator
\begin{equation}
\begin{split}
    & F[h](\theta,\phi)  \\
    & \equiv \alpha \int_0^{\pi} d \theta' \int_0^{2 \pi} d \phi' \sin(\theta') f(\sigma(\theta,\theta', \phi- \phi')) h(\theta',\phi') \label{ret},
\end{split}
\end{equation}
for $f \in [c,z]$ (defined in Eq.~\ref{factor}) corresponding to $F \in [C,Z]$, respectively, and for $h$ any function of $(\theta,\phi)$. $\sigma$ here was already defined in Eq.~\ref{pw}.

Other quantities useful for expressing the steady state will now be defined. The average segment length $l(\theta,\phi)$ in the direction $(\theta,\phi)$ is defined as
\begin{equation}
\begin{split}
	\frac{1}{l(\theta,\phi)}=&-g+d_\text{m}\int d\theta' \int d\phi' \sin(\theta') \left[c (\theta,\theta',\phi-\phi')  \right.
	\\
	&\left. +z(\theta,\theta',\phi-\phi')\right]k(\theta',\phi')
	\label{hlm}.
	\end{split}
\end{equation}
The ratio of inactive (static at both ends) to active (both growing and shrinking) segments in the direction $(\theta,\phi)$ is given by
\begin{equation}
    Q(\theta,\phi)=\frac{m^\text{0}(\theta,\phi)}{m^\text{+}(\theta,\phi)+m^\text{-}(\theta,\phi)} \enspace .
\end{equation}
Using $v^\text{+} m_\text{i}^\text{+} =v^\text{--} m_\text{i}^\text{--}$ in the steady state system with bounded growth, the density of active segments $r(\theta,\phi)$ in the direction $(\theta,\phi)$ is defined as
\begin{equation}
    r(\theta,\phi)=\left(1+\frac{v^\text{+}}{v^\text{--}}\right)l(\theta,\phi)\sum_{\text{i}=1}^{\infty}m_\text{i}^\text{+}(\theta,\phi).
\end{equation}
To formulate non-dimensional steady state equations, a variable $l_{\text{0}}$ is defined with dimensions of length
\begin{equation}
l_{\text{0}}=\left[\left(\frac{1}{v^\text{+}}+\frac{1}{v^\text{--}}\right)\frac{r_{\text{n}}}{4\pi}\right]^{-\frac{1}{4}}, \label{dim}
\end{equation}
which is used to non-dimensionalize parameters and variables. Note that $r_{\text{n}}$ is normalized by $4\pi$ (obtained by integrating the differential solid angle $\sin(\theta) d \theta d \phi$ over all $(\theta,\phi)$) in 3D, in contrast with normalizing by $2\pi$ in 2D.
The dimensionless length ratio $\alpha$, defined by
\begin{equation}
\alpha \equiv d_\text{m} / l_{\text{0}},
\end{equation}
will be an important quantity used to simplify the equations. The dimensionless quantities are then $G \equiv gl_{\text{0}}$, $L \equiv ll_{\text{0}}^{-1}$, $K \equiv kl_{\text{0}}^2$ and $R \equiv rl_{\text{0}}^3$.

The control parameter for this system is therefore the non-dimensional form of $g$ which is explicitly
\begin{equation}
G=\left[\frac{4\pi v^\text{+} v^\text{--} }{r_{\text{n}}(v^\text{+} +v^\text{--})}\right]^{\frac{1}{4}} \left(\frac{r_{\text{r}}}{v^\text{--}}-\frac{r_{\text{c}}}{v^\text{+}}\right). \label{G}
\end{equation}
Using these simplifications and the non-dimensional variables, the steady state equations can be written as
\begin{subequations} \label{four}
\begin{align}
\frac{1}{L(\theta,\phi)}&=-G+C[K](\theta,\phi)+Z[K](\theta,\phi), \label{four1}\\
K(\theta,\phi)&=L(\theta,\phi)(1+Q(\theta,\phi))R(\theta,\phi), \label{four2} \\
Q(\theta,\phi)&=Z[LK(1+Q)](\theta,\phi), \label{four3} \\
R(\theta,\phi)&=L(\theta,\phi)+L(\theta,\phi)K(\theta,\phi)Z[R](\theta,\phi) \label{four4}.
\end{align}
\end{subequations}

There is a symmetry in Eqs.~\ref{four} under $l_{\text{0}} \rightarrow - l_{\text{0}}$ (where $ \alpha \rightarrow - \alpha$ under this transformation). This arises as Eq.~\ref{dim} has a positive and negative root. Since $l_{\text{0}}$ can be either positive or negative and $g<0$ for a physically realizable system, any nonzero real $G \equiv g l_\text{0}$ describes a physically realizable system. Therefore, without loss of generality, throughout this paper, we choose $l_\text{0}<0$ so that $g<0$ corresponds to $G>0$.
\subsubsection{Isotropic Solution}
\label{lw}
Next, the solution of the steady state equations just derived (Eqs.~\ref{four}) will be calculated in the isotropic case. This will use the spherical harmonic functions $Y_{\ell}^\text{m}(\theta,\phi)$ that form the complete set of orthogonal functions defined on the surface of a sphere with two indices ($\ell$ and $\text{m}$). The associated Legendre polynomials $P_\ell^\text{m}$ will also be utilized. Further details of these functions are in Appendix \ref{SphericalHarmonics_Legendre}.

The spherical harmonics defined by Eq.~\ref{sph} provide an orthonormal basis of eigenfunctions of the linear operator defined in Eq.~\ref{ret}, with the eigenvalue equation
\begin{equation}
    F[Y_\ell^\text{m}(\theta,\phi)]=  \frac{4\pi \alpha \mathfrak{F}_\ell}{(2\ell+1)}  Y_\ell^\text{m}(\theta,\phi), \label{eig}
\end{equation}
where a functional redefinition of the form $\mathfrak{F}(cos(\sigma)) \equiv f(\sigma) $ has been introduced, where $f \in [c,z]$ (defined in Eq.~\ref{factor}) corresponds to both $F \in [C,Z]$ and $\mathfrak{F} \in [\mathfrak{C},\mathfrak{Z}]$, respectively, and with $\sigma$ defined in Eq.~\ref{pw}.

To prove Eq.~\ref{eig}, first note that the Legendre polynomials $P_\ell(x)$ defined in Eq.~\ref{lpp} form a complete set of orthogonal functions $P_\ell(x):\mathbb{R}\rightarrow \mathbb{R}$, and a (unique) Legendre expansion is given by
\begin{align}
    &\mathfrak{F}(\cos(\sigma))=\sum_{\ell=0}^{\infty} \mathfrak{F}_\ell P_\ell (\cos(\sigma)),\label{le} \\
    &\text{where} \enspace \mathfrak{F}_\ell=\frac{2\ell+1}{2} \int_{-1}^1  \mathfrak{F}(x)P_\ell(x) dx. \label{stt}
\end{align}
Then, the spherical harmonic addition theorem \cite{Ferrers,SH}
\begin{equation}
    P_\ell(\cos(\sigma))=\frac{4\pi}{2\ell+1}\\ \sum_{\text{m}=-\ell}^\ell Y_\ell^\text{m}(\theta,\phi) {Y^\dag}_\ell^{\text{m}}(\theta',\phi'),
\end{equation}
where $^\dag$ denotes complex conjugation, can be used to rewrite Eq.~\ref{le} as 
\begin{equation}
    f(\sigma)=\sum_{\ell=0}^{\infty} \frac{4\pi}{2\ell+1} \sum_{\text{m}=-\ell}^{\ell} \mathfrak{F}_\ell Y_\ell^\text{m}(\theta,\phi) {Y^\dag}_\ell^{\text{m}}(\theta',\phi').
\end{equation}
Substituting this expression for $f(\sigma)$ into Eq.~\ref{ret} with the spherical harmonic $Y_\ell^\text{m}$ as the argument of $F$ gives
\begin{equation}
\begin{split}
	F[Y_\ell^\text{m}(\theta,\phi)]=  & \alpha \int _0^{\pi} d\theta' \int_0^{2\pi} d\phi' sin(\theta') \sum_{\text{n=0}}^\infty   \sum_{\text{p-=n}}^{\text{n}} \\
	&\frac{4\pi \mathfrak{F}_\text{n}}{2n+1} Y_\text{n}^\text{p}(\theta,\phi) {Y^\dag}_\text{n}^{\text{p}}(\theta',\phi') Y_\ell^\text{m}(\theta',\phi'). \label{simpl}
\end{split}
\end{equation}
Using the standard orthogonality relation of spherical harmonics 
\begin{equation}
\int_0^\pi d\theta' \int_0^{2\pi} d\phi' \sin(\theta') Y_\ell^\text{m}(\theta',\phi')Y_{\ell'}^{\text{m}'}(\theta',\phi')=\delta_{\text{ll'}}\delta_{\text{mm'}}, \label{orth}
\end{equation}
where the Kronecker delta $\delta_{\text{ab}}$ takes the value 1 iff $\text{a} \equiv \text{b}$ and 0 otherwise, simplifies Eq.~\ref{simpl} to
\begin{equation}
    F[Y_\ell^\text{m}(\theta,\phi)]=\sum_{\text{n=0}}^\infty   \sum_{\text{p=-n}}^{\text{n}}   \frac{4\pi \alpha \mathfrak{F}_\text{n}}{2n+1} Y_\text{n}^\text{p}(\theta,\phi) \delta_{\text{nl}} \delta_{\text{pm}},
\end{equation}
which is equivalent to Eq.~\ref{eig}.

Using this, the isotropic solution can now be calculated. In the isotropic (and stationary) state of the system, which will be denoted by overbars, all angular dependence drops out and Eqs.~\ref{four} becomes
\begin{subequations} \label{iso}
\begin{align}
\frac{1}{\bar{L}}&=-G+4 \pi \alpha (\mathfrak{C}_\text{0} +  \mathfrak{Z}_\text{0}) \bar{K}  \label{iso1} \\
\bar{K}&=\bar{L}(1+\bar{Q})\bar{R}  \label{iso2}  \\
\bar{Q}&=4\pi \alpha \mathfrak{Z}_\text{0}\bar{L}\bar{K}(1+\bar{Q})   \label{iso3} \\
\bar{R}&=\bar{L}+4\pi \alpha \mathfrak{Z}_\text{0} \bar{L}\bar{K}\bar{R} \label{iso4}.
\end{align}
\end{subequations}
where the identity $F[1]=4\pi \alpha \mathfrak{F}_\text{0}$ was used, which arises from setting $\ell=\text{m}=0$ in Eq.~\ref{eig}.

Substituting and rearranging Eqs.~\ref{iso} reduces to the expression
\begin{equation}
    \bar{K}(4\pi \alpha \mathfrak{C}_\text{0} \bar{K}-G)^2=1 \label{hold},
\end{equation}
with a full derivation given in Appendix~\ref{calc3}. The quantity $\bar{K}$ is always positive. $\mathfrak{C}_\text{0}$ is also taken to be positive and it will be shown in Sec.~\ref{subsection:positivec} that this agrees well with experimental values. If $\alpha<0$ and $G>0$, then Eq.~\ref{hold} gives an expression for $G$ in terms of $\bar{K}$ as
\begin{equation}
G=4\pi \alpha \mathfrak{C}_\text{0} \bar{K} +\bar{K}^{-\frac{1}{2}}. \label{monotonic}
\end{equation}
For given values of $\alpha<0$ and $\mathfrak{C}_\text{0}$, there is clearly a uniquely determined value of $G$ for each value of $\bar{K}$. The converse holds since the right hand side of Eq.~\ref{monotonic} is a strictly decreasing function of positive $\bar{K}$ which tends to $\infty$ for $\bar{K} \rightarrow 0$ and tends to $-\infty$ for $\bar{K} \rightarrow \infty$. Furthermore, $G>0$ gives us the constraint $\bar{K}< |4\pi \alpha \mathfrak{C}_\text{0} |^{-2/3}$ (the value of $\bar{K}$ for which $G=0$).

An identical argument for $G<0$, $\alpha>0$ leads to the same conclusions, giving the general constraint that holds for both the cases $(\alpha<0$, $G>0)$ and $(\alpha>0$, $G<0)$ as
\begin{equation}
    0<\bar{K}< |4\pi \alpha \mathfrak{C}_\text{0} |^{-\frac{2}{3}}.
\end{equation}
Therefore, there is a density limit to the system above which it is not possible to have an isotropic system. This density limit decreases with increasing $d_\text{m} $ (which is equivalent to increasing $|\alpha|$). This makes sense physically since increasing $d_\text{m}$ increases the range of interaction between segments, therefore preventing an isotropic solution at lower density systems.

Additionally, as it will be needed in Sec.~\ref{subsection:positivec}, note that Eqs. \ref{iso} also lead to to the expression
\begin{equation}
    G=\frac{4 \pi \alpha (\mathfrak{C}_\text{0} + \mathfrak{Z}_\text{0}) \bar{N} -1 }{\bar{N}^{\frac{1}{3}} (4 \pi \alpha \mathfrak{Z}_\text{0} \bar{N} -1)^{\frac{2}{3}}} \label{cubed},
\end{equation}
where $N\equiv LK$, with the detailed derivation provided in Appendix~\ref{calc3}.
\subsection{Simulation}
\label{subsection:simulation}
The predictions from the 3D mean-field theory mathematical model just derived will later be validated against 3D microtubule simulations using Tubulaton \cite{tubulatonwebsite}, which will be briefly explained here. Tubulaton uses a discretized model of microtubule dynamics modeling each microtubule individually as a line of end-to-end unit vectors, each corresponding to $8$nm and representing a single ring of tubulin. Microtubules grow and shrink via the addition or removal of unit vectors at their ends. Microtubule interaction dynamics such as zippering and induced catastrophe (Fig.~\ref{drawing}) are incorporated, as well as individual microtubule dynamic behaviors such as nucleation and spontaneous catastrophe. An external membrane is prescribed within which microtubules remain. Previous papers have provided a detailed description of Tubulaton \cite{tubulaton,durand}, so here we focus on specific changes and additions that have been made with the purpose of comparing to the mean-field theory model.

To reflect mean-field theory defined within an infinite volume without a boundary, we construct three spheres of decreasing radii all centred on the same point. This setup was chosen instead of periodic boundary conditions to avoid complications arising from how microtubules would interact with each other at the boundary. The largest sphere forms the external boundary of the system, the middle sphere defines the region where microtubules nucleate randomly (both spatially and directionally), and the smallest sphere is the region within which we calculate the level of anisotropy.

Tubulaton was extended to improve the originally encoded assumption that the probabilities of induced catastrophe and zippering are fixed above and below a threshold angle (normally prescribed as $40$ degrees). For this work, they are varied as a function of collision angle. This reflects experimental observations \cite{angles} and the mean-field assumptions. For our simulations including the effects of zippering we match experimental observations \cite{angles} (described later in detail and illustrated by Fig.~\ref{angle}), or we set it equal to zero for all angles to remove the effects of zippering.

The ability to vary $r_{\text{c}}$ and $r_{\text{n}}$ was already incorporated into Tubulaton. Effects of spontaneous rescue are not included but these are unnecessary since the regime of interest is negative $g$ (noting that $G$ can be either positive or negative) and varying $r_{\text{c}}$ whilst $r_r=0$ in Eq.~\ref{defcp} gives the full range of negative values of $g$, which is still the case when we take $v^\text{+}=v^\text{--}$ for all our simulations. However, the region of values $(r_\text{c},r_\text{n})$ within which we performed simulations was restricted by convergence and computational time. At low $r_{\text{c}}$, we observed slower convergence and very large fluctuations in segment density for small changes in $r_{\text{n}}$. On the other hand, at large $r_{\text{c}}$, we observed that the value of $r_{\text{n}}$ required for a steady state density grew sharply leading to long computational times.

\begin{figure}
\vspace{0cm}
\hspace{0cm}
    \captionsetup{justification=raggedright}
    \includegraphics[width=8cm]{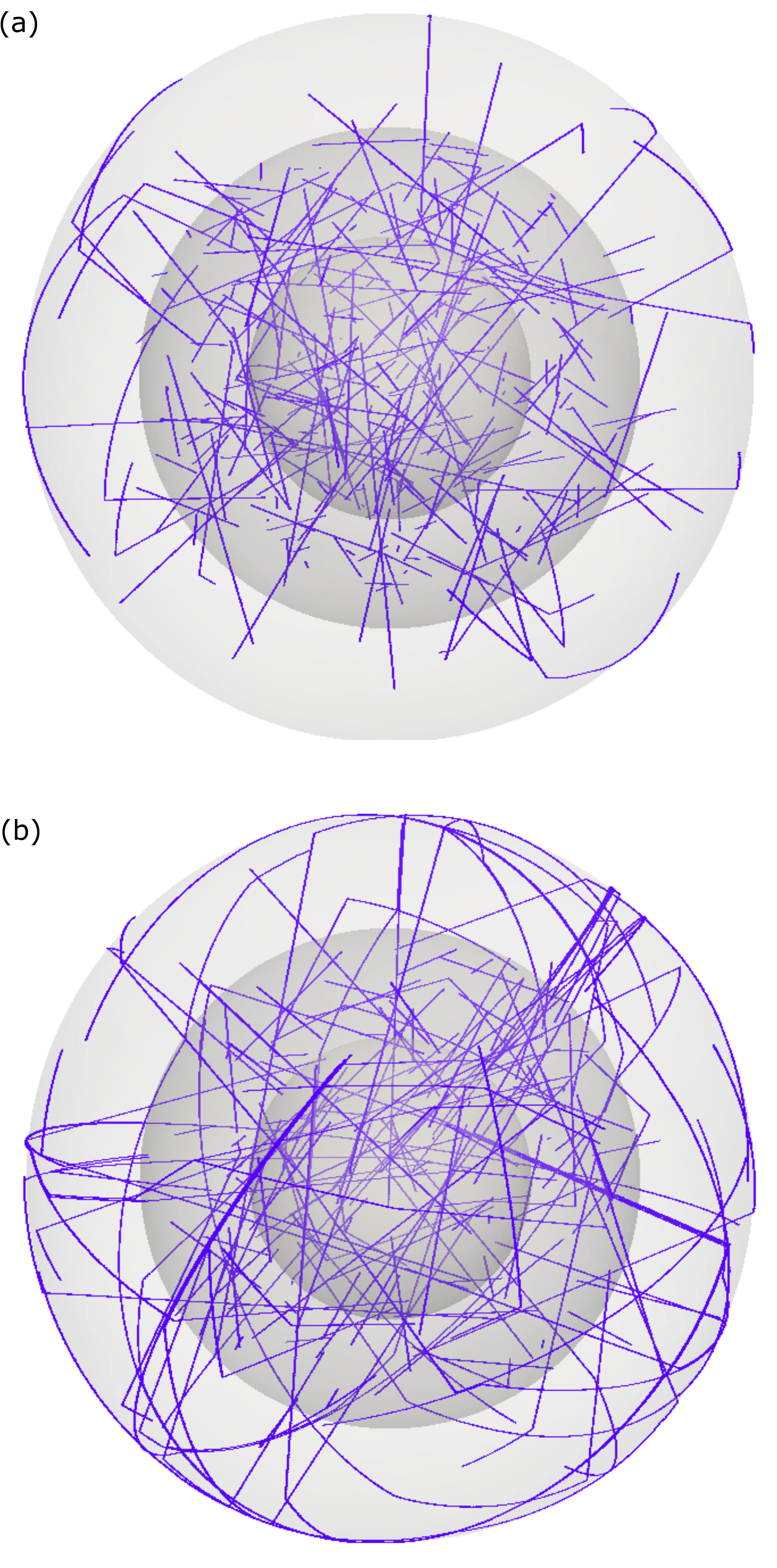}
    \caption{Snapshot from two Tubulaton simulations showing the microtubules and sphere boundaries. The three shaded spheres represent the external membrane (1000 unit radius), nucleation region (700 unit radius) and anisotropy analysis region (400 unit radius). Simulations are with zippering where (a) $r_{\text{c}}=3.5 \times 10^{-3}$, $r_{\text{n}}=4 \times 10^{-10}$ (b) $r_\text{c}=10^{-3}$, $r_{\text{n}}=6 \times 10^{-11}$. Other important parameter values can be found in Table~\ref{ParamTable}.}
    \label{pap}
\end{figure}

Details of the parameter values used in the simulations are included in Appendix~\ref{AppendixD} and shown in Table~\ref{ParamTable}. Within the limit of $10,000$ time steps ($15$-$30$ minutes of simulated time), the number of microtubule segments converged (Fig. \ref{sim}) so simulations are expected to have reached a steady state as was assumed in the mean-field theory analysis. Snapshots of two different Tubulaton simulations are illustrated in Fig.~\ref{pap}.
%
\subsection{Order Parameter}
\label{ss:order}
Following earlier comparisons to nematic liquid crystals, the standard nematic order parameter \cite{liqc,liqcpedagog,LAMY} is used to quantify the alignment of the microtubules in the system. In $D$ dimensions, this is a unique tensor up to an overall factor, which is normalized by setting it equal to unity in the completely anisotropic state. It is defined as
\begin{equation}
{S}_{\text{ab}}=\left\langle \frac{D}{D-1}n_\text{a} n_\text{b} -\frac{1}{D-1}\delta_{\text{ab}} \right\rangle, \label{generalOP}
\end{equation}
where $\langle \enspace \rangle$ denotes taking a weighted average over every microtubule segment, each of which is parameterized by the $D$-dimensional unit vector $n$ and weighted by the segment length. Performing this averaging using an integral weighted by the microtubule length density in each direction (with $D=3$), gives the $3\text{x}3$ matrix with components (with $i,j=1,2,3$).
\begin{equation}
S_{ij}= \frac{\int_0^{\pi}d\theta \int_0^{2\pi} d\phi \sin(\theta) K(\theta,\phi) \left(\frac{3}{2}n \otimes n
-\frac{1}{2} \mathcal{I}\right)_{ij}}{\int_0^{\pi}d\theta \int_{0}^{2\pi} d\phi \sin(\theta) K(\theta,\phi)}. \label{op}
\end{equation}
where $\otimes$ is the outer product. For calculating the order parameter $S$ in simulations, the discretized form of Eq.~\ref{op} is used
\begin{equation}
S_{ij}= \frac{ \sum_\mu \left(\frac{3}{2}n^\mu \otimes n^\mu -\frac{1}{2} \mathcal{I}\right)_{i j}}{\sum_\mu 1}, \label{opp}
\end{equation}
where $\mu$ labels each microtubule segment, each of which is associated to a unit direction vector $n^\mu$.

This matrix with components $S_{ij}$ has three eigenvalues which are all zero if and only if the system is in its completely isotropic state. We define $S$ as the maximal absolute value of the eigenvalue of this matrix. A larger value of $S$ indicates a higher level of anisotropy so an increasing $S$ indicates that a system is changing from an isotropic state to an anisotropic state \cite{orderparam}.
\subsection{Visualization and code availability}
All numerical work and graphing in Sec.~\ref{section:section_two} was performed using MATLAB R2021a or Python. Simulations were visualized using Paraview 5.0.1. All scripts used for running and plotting outputs from Tubulaton simulations are available in the software repository at \url{https://gitlab.com/slcu/teamHJ/publications/gibson_etal_2023}, and Tubulaton is available at \url{https://gitlab.com/slcu/teamhj/tubulaton}.
\section{Results}
\label{section:section_two}
\subsection{Conditions for a Phase Transition}
\label{subsection:twosix}
Here, we consider constraints on microtubule properties which allow for a change from disorder to order as $g$ is increased in order to determine the conditions for a phase transition to occur. Assuming that the change in order is continuous, when the order parameter transitions from zero to small and non-zero, the steady-state solution will be a small perturbation from the isotropic solution. Therefore, we will perturb the isotropic solution from Sec. \ref{lw} and determine what conditions allow a solution to exist with small order parameter. The perturbative solution of Eqs.~\ref{four} is derived in Appendix~\ref{calc2} and results in the eigenvalue equation
\begin{equation}
\left(1- 4\pi \alpha \mathfrak{Z}_\text{0}\bar{N} \right)\kappa(\theta,\phi)=-2\bar{N}C[\kappa(\theta,\phi)] \label{1st},
\end{equation}
where $\bar{N}=\bar{L} \bar{K}$ and $K=\bar{K}(1+\kappa)$ defines the first order perturbation $\kappa(\theta,\phi)$. This eigenvalue equation is a special case of Eq.~\ref{eig} so the spherical harmonics defined in Eq.~\ref{sph} form an orthonormal basis of eigenfunctions. Since the eigenvalues in Eq.~\ref{eig} depend only on the lower index $\ell$ of the spherical harmonics, each $\ell$ determines a different value of $\bar{N}$, each denoted by $\bar{N}^*_\ell$, given explicitly by
\begin{equation}
\bar{N}^*_\ell=\left(4\pi \alpha \mathfrak{Z}_\text{0}-\frac{8\pi \alpha}{2\ell+1}\mathfrak{C}_\ell \right)^{-1},
\end{equation} 
for which there is a potential phase transition.
Directly substituting these values of $\bar{N}$ into Eq.~\ref{cubed} gives us the corresponding values of $G$ as
\begin{equation}
G^*_\ell=\left[1+\frac{(2l+1)\mathfrak{C}_\text{0}}{2\mathfrak{C}_\ell}\right]\left[-\frac{ 8\pi \alpha \mathfrak{C}_\ell}{2l+1}\right]^{\frac{1}{3}}\label{GG},
\end{equation}
which by extension are also indexed by $\ell$.

The location of the possible phase transition depends only on the catastrophe probability function and is independent of the zippering probability function since there is no dependence on $\mathfrak{Z}_\ell$ in Eq.~\ref{GG}. This is similar to the 2D case \cite{main}.
\begin{figure}
\vspace{-.5cm}
\hspace{0cm}
    \captionsetup{skip=0.2cm,margin={0cm,0cm},justification=raggedright}
    \includegraphics[width=1\linewidth,height=7cm]{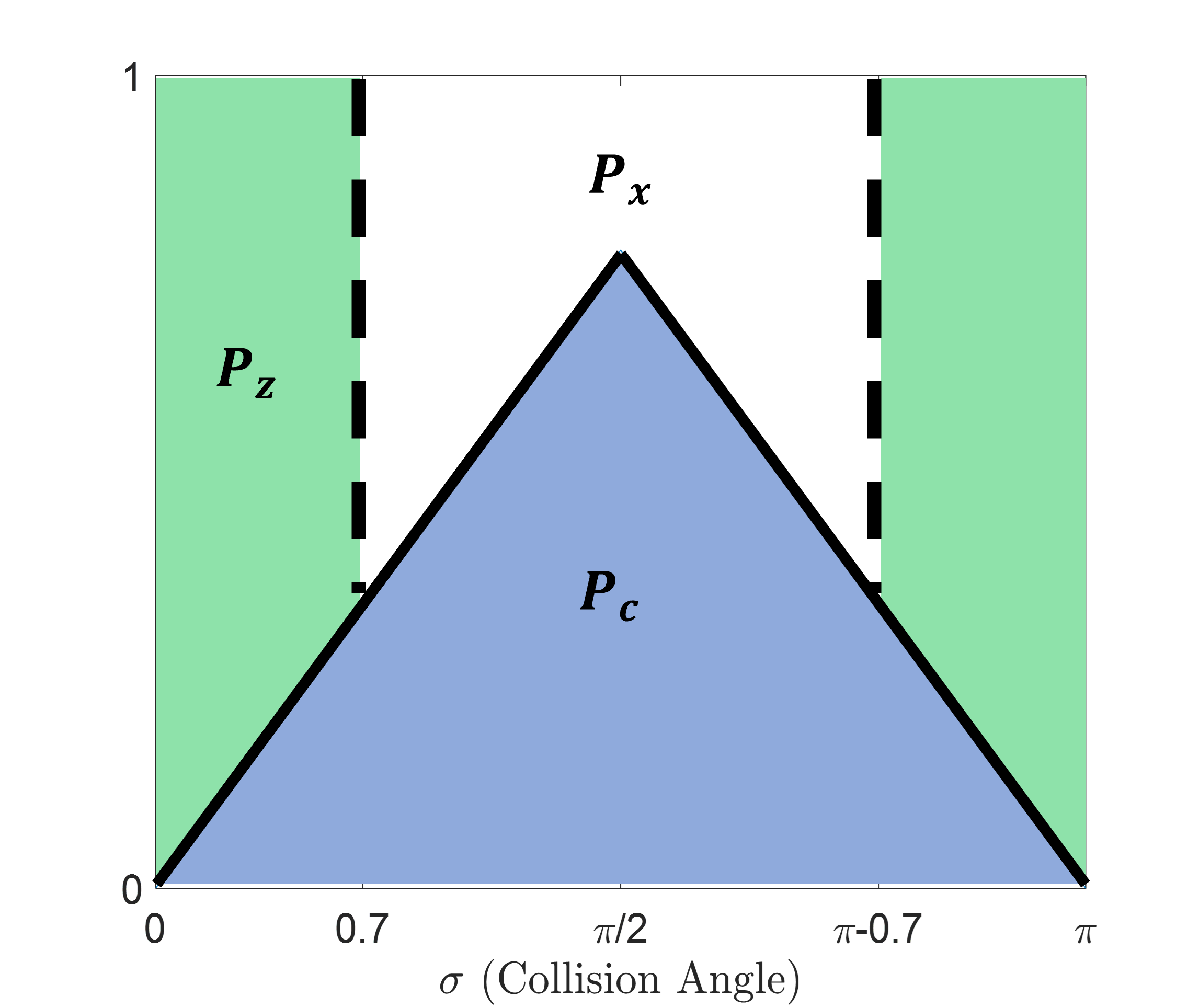}
    \caption{Probability of induced catastrophe ($P_\text{c}$), induced zippering ($P_\text{z}$) and crossover ($P_\text{x}=1-P_\text{c}-P_\text{z}$) as a function of collision angle $\sigma$ used in our analytical calculation and computational simulations reflecting experimental observations. The point of maximum $P_\text{c}$ is at $(\pi/2,\pi/4)$.}
    \label{angle}
\end{figure}
\addtocounter{figure}{-1}
\begin{figure*}
\hspace{-.9cm}
\begin{subfigure}[t]{0.03\textwidth}
\vspace{-3.2cm}
    \text{(a)}
  \end{subfigure}
\begin{subfigure}{.4\textwidth}
\addtocounter{figure}{1}
\vspace{0cm}
\captionsetup{justification=raggedright,skip=0cm}
\includegraphics[clip,width=1.1\textwidth,height=7cm]{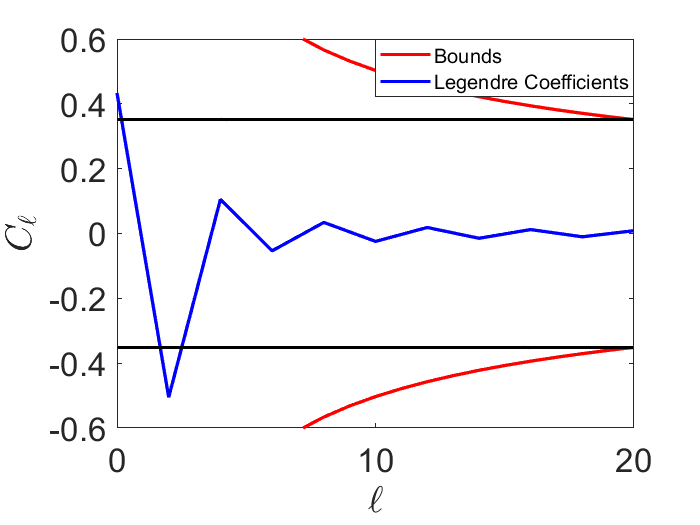}
\phantomcaption
\label{fig_1}
\end{subfigure}
\hspace{1.5cm}
\begin{subfigure}[t]{0.03\textwidth}
\vspace{-3.2cm}
    \text{(b)}
  \end{subfigure}
\begin{subfigure}{.4\textwidth}
\vspace{0cm}
\captionsetup{justification=raggedright,skip=0cm,margin={.5cm,1cm}}
\includegraphics[width=1.1\textwidth,height=7cm]{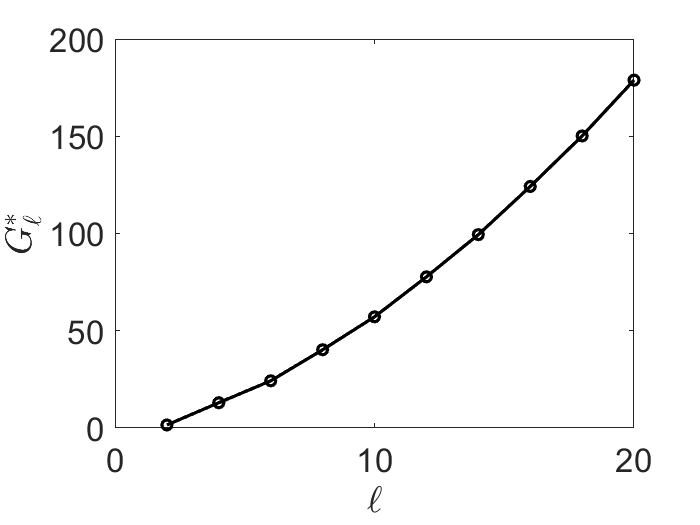}
\phantomcaption
\label{fig_2}
\end{subfigure}
\vspace{0cm}
\captionsetup{skip=0cm,justification=raggedright}
\label{double}
\vspace{0cm}
\captionsetup{skip=0cm,justification=raggedright}
\caption{Effect of varying the Legendre coefficient index $\ell$ (for even values of $\ell$ between $0$ and $20$) of the spherical harmonic perturbation $\kappa=Y_{\ell}^\text{m}$ (normalized as in Eq.~\ref{sph}) on (a) Legendre coefficients $\mathfrak{C}_\ell$ as defined in Eq.~\ref{graphlink} and (b) corresponding control parameter values $G_{\ell}^*$. In (a) the red line shows the analytical upper and lower bounds of the Legendre coefficients while the black horizontal lines are equal to the size of these bounds at $\ell=20$ and is included for comparison.}
\end{figure*}
In order to explicitly evaluate Eq.~\ref{GG}, we approximate experimental observations that catastrophe probability increases approximately linearly as a function of collision angle to around 0.7 at $\pi/2$ collision angle \cite{angles}. $P_\text{c}(\sigma)$ is therefore chosen to be two concatenated linear functions through the three points $(0,0)$, $(\frac{\pi}{2},\frac{\pi}{4})$, and $(\pi,0)$ (Fig.~\ref{angle}). Zippering probability is similarly approximated, with zippering occurring primarily at lower angles, by taking $P_\text{z}(\sigma)=1-P_\text{c}(\sigma) \enspace \forall  \sigma \leq 0.7  \text{, } \sigma \geq \pi-0.7$ and $P_\text{z}(\sigma)=0$ otherwise (Fig.~\ref{angle}). 

Quantities required to explicitly calculate the possible phase transition location can now be numerically calculated. In particular,
\begin{equation}
    \mathfrak{C}_\ell=\frac{2\ell+1}{2} \int_0^{\pi} \sin^2(y) P_\text{c}(y) \\P_\ell(\cos(y)) dy, \label{graphlink}
\end{equation}
an expression obtained from Eq.~\ref{stt} through the change of variables $x \equiv \cos(y)$. Splitting the integral's domain in half, making a change of variables in $[0,\pi/2]$ by $y \rightarrow (\pi-y)$ and using the identity $P_\ell(-x)=(-1)^\ell P_\ell(x)$, results in
\begin{equation}
\mathfrak{C}_\ell=\frac{\left(1+(-1)^\ell\right)(2\ell+1)}{2} \int_0^{\frac{\pi}{2}} \frac{1}{2} y \sin^2(y)  P_\ell(\cos(y)) dy.
\end{equation}

The Legendre coefficients $\mathfrak{C}_\ell$ are fundamental to evaluating the values of the control parameter which give rise to a potential phase transition. $\mathfrak{C}_\ell$ identically vanishes for all odd $\ell$. The values of $\mathfrak{C}_\ell$ for even $\ell$ up to $\ell=20$ are plotted in Fig.~\ref{fig_1}, whose magnitudes decrease in this range. Note that $\mathfrak{C}_\text{0}=0.43$ (2sf) provides a positive value for $\mathfrak{C}_\text{0}$ which was previously asserted as reasonable for Eq.~\ref{hold}. As shown in \cite{coef} using an improved Bernstein inequality \cite{antonov,lorch}, there is a general bound on Legendre coefficients
\begin{equation}
    \left|\mathfrak{C}_\ell \right| \leq \frac{2}{\sqrt{\pi(2\ell-1)}} \int_{-1}^1 \frac{|\mathfrak{C}'(x)|}{(1-x^2)^{\frac{1}{4}}} dx, \label{bounds}
\end{equation}
which is plotted on Fig.~\ref{fig_1}. Since the magnitude of the bound decreases as $\ell$ increases, the bound at $\ell=20$ shows that $\mathfrak{C}_2$ must be the largest Legendre coefficient. 

A unique value of the control parameter $G^*_\ell$ corresponds to each Legendre coefficient (Fig.~\ref{fig_2}) for even $\ell$ up to $\ell=20$ for which the value $G^*_\ell$ (Eq.~\ref{GG}) increases monotonically. Since all values of $G^*_\ell$ are positive, $l_{\text{0}}$ can be taken to be negative so that $g=G/l_{\text{0}}$ is negative since we only wish to consider physically realizable solutions with bounded growth (as discussed in Sec.~\ref{subsection:cp}). However, no particular value of $G^*_\ell$ has been singled out to correspond to a phase transition. To do this, it is necessary to consider which perturbations ($\ell$) leads to a change in the order parameter defined in Eq.~\ref{op}. 

The perturbed order parameter can be calculated directly. Perturbing away from the isotropic case $K=\bar{K}$, where the order parameter is zero, the new variable $K \equiv \bar{K} \left(1+\sum_{\ell,\text{m}} \beta_\ell^\text{m} Y_\ell^\text{m} \right)$ is defined, where $\beta$ are taken to be small constants. Expanding Eq.~\ref{op} to first order in $\beta$, the order parameter for the perturbed system is
\begin{equation}
    S_\beta=\sum_{\ell=\text{0}}^{\infty} \sum_{\text{m}=-\ell}^\ell \beta_{\ell}^\text{m} \frac{\int_0^\pi d \theta \int_0^{2\pi} d \phi \sin(\theta)  Y_\ell^\text{m} (\frac{3}{2}
     n \otimes n - \frac{1}{2} \mathcal{I})}{\int_0^\pi d \theta \int_0^{2\pi} d \phi \sin(\theta)  } \label{per}.
\end{equation}
The tensor components $Q_{\text{ij}} \equiv \left(\frac{3}{2} n \otimes n -\frac{1}{2} \mathcal{I}\right)_{\text{ij}}$ from Eq.~\ref{generalOP} can be written in terms of only $\ell=2$ spherical harmonics as
\begin{align}
    Q_\text{11} & =  \sqrt{\frac{3 \pi} { 10}} \left(Y_\text{2}^\text{-2} + Y_\text{2}^\text{2} \right) - \sqrt{\frac{\pi}{5}}Y_\text{2}^\text{0} ,   \nonumber\\
    Q_\text{22} & =-  \sqrt{\frac{3 \pi} { 10}} \left(Y_\text{2}^\text{-2} + Y_\text{2}^\text{2} \right)- \sqrt{\frac{\pi}{5}}Y_\text{2}^\text{0} , \nonumber \\
    Q_\text{33} & =  \sqrt{\frac{4\pi}{5}} Y_\text{2}^\text{0}, \label{q} \\
    Q_\text{12} & =  \sqrt{\frac{-3 \pi}{10}} \left(Y_\text{2}^\text{-2}-Y_\text{2}^\text{2} \right), \nonumber \\
    Q_\text{13} & = \sqrt{\frac{3\pi}{10}} \left(Y_\text{2}^\text{-1}-Y_\text{2}^\text{1}\right), \nonumber \\
    Q_\text{23} & =\sqrt{\frac{-3\pi}{10}} \left(Y_\text{2}^\text{-1}+Y_\text{2}^\text{1}\right). \nonumber
\end{align}
Due to the spherical harmonic orthogonality condition of Eq.~\ref{orth}, the only nonzero contribution to Eq.~\ref{per} is from $\beta_2^\text{m}$ for $\text{m}=-2,-1,0,1,2$. This claim was verified numerically by calculating the eigenvalues of the order parameter for perturbations $\kappa=\beta Y_\ell^\text{m}$ for $\ell=1,...,200$ (testing each of $\text{m}=-2\ell,...,2\ell$ in turn) for $\beta=1$. All $\ell \neq 2$ perturbations led to zero eigenvalues. The eigenvalues of the order parameter $S$ were $(0.77,-0.77,0)$ for $\kappa =Y_2^\text{m}$ for $\text{m} =-2,-1,1,2$ and $(-0.06,-0.06,0.13)$ for $\text{m}=0$. This confirms that only the $\ell=2$ spherical harmonic perturbation causes disorder in the system.

Therefore, setting $l=2$ in Eq.~\ref{GG}, this mean-field theory model can only exhibit a phase transition at $G^*\equiv G^*_2=-1.56 \alpha^{1/3}$.

\subsection{Comparison to a 2D Model}
\label{subsection:positivec}
The 3D model presented here is based on but is different to the earlier 2D model \cite{main}, and here we highlight three of the main differences.

First, microtubule collision is different in 3D compared to 2D, as two infinite non-parallel lines will always intersect in 2D but not in 3D, so in 3D microtubule thickness plays a more important role in determining the conditions for a potential phase transition. Mathematically, this causes the induced catastrophe flux term (Eq.~\ref{IC}) to be different in 2D and 3D. In 3D, a new factor of $d_\text{m}$ is necessary for dimensional agreement and for the effect of microtubule thickness on collision probability to be reflected in the flux term. Additionally, the factor $|\sin(\sigma)|$ which adjusts for varying collision probability on the angle has a different angular dependence with the collision angle defined in 3D (Eq.~\ref{pw}) as $\sigma=\arccos(\sin(\theta)\sin(\theta')\cos(\phi-\phi')+\cos(\theta)  \cos(\theta'))$ instead of $\sigma=(\phi-\phi')$ in 2D \cite{main} with $\phi$ defined as the 2D polar angle in the usual way.

Second, a technical difference is the role that spherical harmonic modes have taken in this 3D mean-field theory, replacing a similar role played by Fourier modes in 2D. The change is caused by the factor of $|\sin(\sigma)|$ in the catastrophe flux term, which significantly changes the calculation in Sec.~\ref{lw}, and leads to a different linear operator defined in Eq.~\ref{ret}. In 3D, the spherical harmonics provide an orthonormal basis of eigenfunctions, instead of Fourier modes in 2D. Viewing the spherical harmonics $Y_\ell^\text{m}:S^\text{2}\rightarrow \mathbb{R}$ as a higher dimensional analog of $S^\text{1} \rightarrow \mathbb{R}$ Fourier modes \cite{ddim}, this higher dimensional generalization is not unexpected. As a result of this difference, Eq.~\ref{iso} and Eq.~\ref{hold} both involve Legendre coefficients in 3D instead of Fourier coefficients in 2D.

Third, the predicted value of $G$ for which a phase transition is possible takes a different form in 3D compared to 2D, although there are subtle similarities. Specifically, $G \propto l_\text{0}$ and in 3D $l_{\text{0}}$ itself is a fourth root (Eq.~\ref{dim}), so taking the negative solution for $l_\text{0}$ results in positive $G=gl_{\text{0}}$ when $g$ is negative, which is a requirement for bounded growth. The non-dimensional control parameter $G$ has a different form in 3D and 2D, despite the dimensional control parameter $g=r_{\text{c}}/v^\text{--} -r_{\text{r}}/v^\text{+}$ remaining the same. This arises as there are expected differences in the definition of $l_{\text{0}}$, necessary to ensure it has dimensions of length. There are also differences in the factors of $l_{\text{0}}$ involved in non-dimensionalizing $g$ (to $G$) and other variables describing the system. The control parameter can be modified by multiplying by any dimensionless function without fundamentally changing the system, but the expression for the critical value must be correspondingly updated. Since a dimensionless factor of $\alpha^{1/3}$ enters the expression for the critical value of $G$ (Eq.~\ref{GG}) at which a phase transition is possible, we define an effective control parameter $G_{\text{eff}}\equiv - G \alpha^{-1/3}= -g l_{\text{0}}^{4/3}d_\text{m}^{-1/3}$. Interestingly, written in this form, this 3D effective control parameter is proportional to the 2D control parameter in \cite{main} multiplied by a factor of $d_\text{m}^{-1/3}$. Since our prediction for $G_\text{eff}$ is a purely numerical quantity, due to the critical value $G^*\equiv -1.56 \alpha^{1/3}$ now being equivalent to $G_\text{eff}^*=1.56$, this will be used when we compare to simulation in the next section. $G_\text{eff}$ can be written explicitly as
\begin{equation}
    G_\text{eff}=-\left[\frac{4\pi v^\text{+} v^\text{--} }{r_{\text{n}}(v^\text{+} +v^\text{--})}\right]^{\frac{1}{3}} \left(\frac{r_{\text{r}}}{v^\text{--}}-\frac{r_{\text{c}}}{v^\text{+}}\right) d_\text{m}^{-\frac{1}{3}}.
    \label{geff}
\end{equation}
\subsection{Comparison to Simulation}
\label{subsection:compare}
\begin{figure}
\vspace{-.8cm}
\hspace{-.5cm}
    \includegraphics[width=8cm]{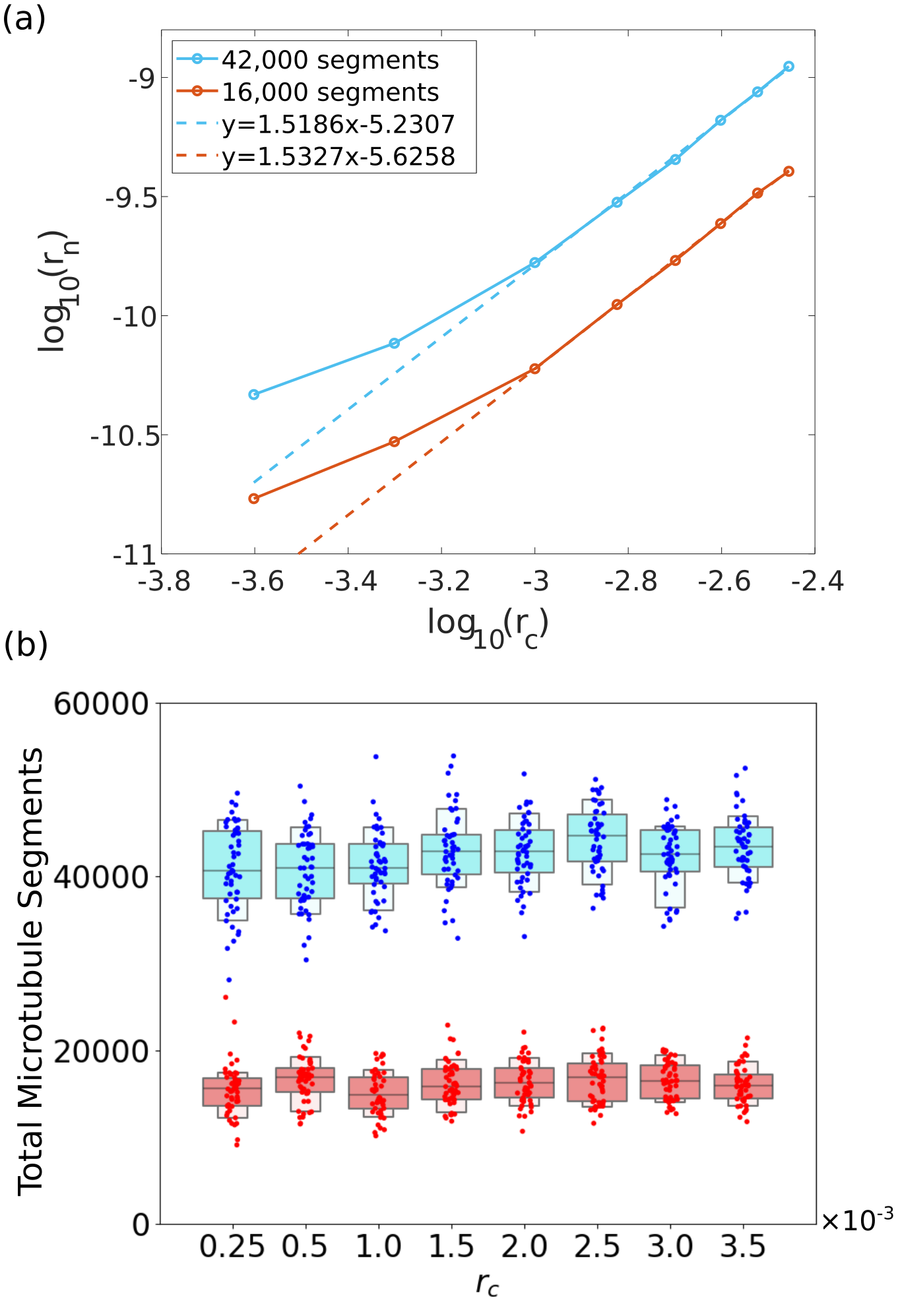}
\captionsetup{justification=raggedright,belowskip=-.4cm}
    \caption{Paths of constant density in the $r_{\text{n}}-r_{\text{c}}$ plane. (a) Paths of approximately constant density for two specified segment numbers. (b) Associated number of segments obtained from 50 simulations for each parameter pair from the two paths in (a). Simulation parameter values are provided in Table~\ref{ParamTable}.}
    \label{fig_5}
\end{figure}
Next, we compare our mean-field theory predictions against computational simulations (described in Sec.~\ref{subsection:simulation}). Specifically, we test three mean-field theory predictions: decreasing $G_\text{eff}$ causes an increase in anisotropy; a phase transition is only possible at approximately $G_\text{eff}^*=1.56$ (3sf); and the zippering probability function does not affect the other two predictions.

From Eq.~\ref{geff}, $G_{\text{eff}} \propto r_\text{n}^{-1/3}$ and $G_{\text{eff}} \propto r_\text{c}$ so to vary $G_{\text{eff}}$ in simulations, we directly vary $r_{\text{n}}$ and $r_{\text{c}}$. We make a specific choices of $r_{\text{n}} \text{ and }r_{\text{c}}$ to keep the average steady state number of microtubule segments in the volume within which we perform our anisotropy calculation within a fixed range (Fig.~\ref{fig_5}). This is to ensure that microtubule density is not affecting the anisotropy levels of the system. A dimensional argument determines the relationship between $r_\text{c}$ and $r_\text{n}$ required to keep the microtubule density constant. Microtubule lifetime is proportional to $1/r_{\text{c}}$ (this is clearer in the case without rescue), resulting in a mean microtubule length proportional to $v^\text{+}/r_{\text{c}}$. The number of microtubules is proportional to $r_{\text{n}}/r_{\text{c}}$ for similar reasons. Therefore, imposing that the total density of microtubule segments be constant is equivalent to the condition that $v^\text{+}/r_{\text{c}} \times r_{\text{n}}/r_{\text{c}}$ be constant. This suggests that when $v^\text{+}$ is kept constant, $r_{\text{n}} \propto r_{\text{c}}^2$ ensures a constant density of microtubule segments. We test this prediction in simulations within a $(r_\text{n},r_\text{c})$ region where density fluctuations (after converging to a steady state density) were small and convergence occurred within reasonable computation times (Fig.~\ref{fig_5}). The $(r_\text{n},r_\text{c})$ path of constant microtubule density for approximately $42000$ and $16000$ segments (Fig.~\ref{fig_5}b) both give an approximate power law $r_{\text{n}} \propto r_{\text{c}}^{1.5}$ at larger values of $r_\text{c}$ (Fig.~\ref{fig_5}a) which is similar to our predicted quadratic power law in the completely isotropic regime (corresponding to larger $r_{\text{c}}$ or equivalently higher $G_\text{eff}$). Differences between the simplistic theoretical prediction and computational result could be explained, for example, by not being in the isotropic limit in the computational simulations or interactions not being included within the theoretical prediction. Theoretically, our mean-field theory model predicts an increase in anisotropy for lower $r_{\text{c}}$ (corresponding to lower $G_\text{eff}$) which matches where the theoretically predicted power law for the completely isotropic system is seen to break down (Fig.~\ref{fig_5}a).
\begin{figure}
\qquad \qquad \qquad
\begin{minipage}{.48\textwidth}
\vspace{-.7cm}
\hspace{-1cm}
\captionsetup{skip=.5cm,justification=raggedright,margin={0cm,0cm},belowskip=.4cm}
\includegraphics[width=7.8cm]{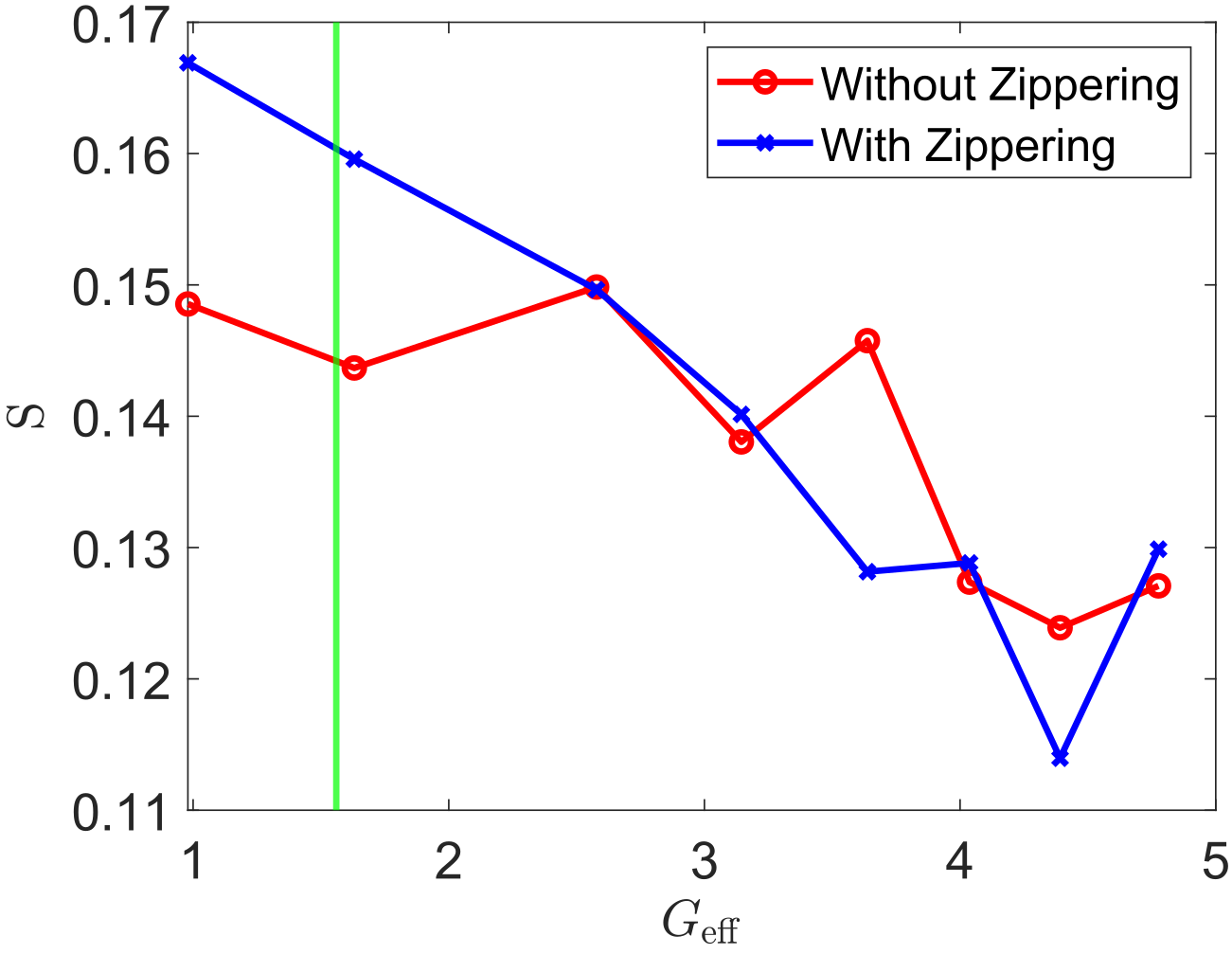}
\caption{Effect of varying effective control parameter $G_{\text{eff}}$ on mean anisotropy $S$ with (blue) and without (red) zippering. Each point is averaged over 50 simulations. The green line marks the mean-field theory prediction of the only place a phase transition can occur at $G_\text{eff}^*=1.56$. Simulation parameter values are provided in Table~\ref{ParamTable}.}
\label{fig_6}
\end{minipage}
\vspace{-.5cm}
\end{figure}
We now investigate how the anisotropy of the system changes with varying $G_\text{eff}$. The mean-field theory prediction that increasing $G_{\text{eff}}$ corresponds to decreasing mean anisotropy (estimated by $S$), indicating a change from a more ordered to less ordered state, is confirmed in the simulations (Fig.~\ref{fig_6}). Furthermore, the $G_{\text{eff}}$ region for which we are observing this decrease in microtubule order supports the prediction of $G^*_{\text{eff}}=1.56$ as an order of magnitude estimate for a change from an isotropic to anisotropic system. 

Next, we investigate the effect of zippering on the change in anisotropy for varying $G_{\text{eff}}$. The similar behavior with and without zippering verifies the prediction that changing $P_\text{z}(\sigma)$ has affects neither the orientational dynamics of the system (Fig.~\ref{fig_6}) nor the density (Fig.~\ref{sim}d), the second of these being important for a fair comparison. Comparing the simulations at the lowest and highest values of $G_{\text{eff}}$ in Fig.~\ref{fig_6} for the zippering and no-zippering case we observe a statistically significant decrease in anisotropy in both cases (using a 2-sample Kolmogorov-Smirnoff test, p-values $4.89 \times 10^{-4}$ and $1.98 \times 10^{-2}$, respectively). This is despite stochastic differences (Appendix~\ref{AppendixC}) in anisotropy between simulations at each $G_{\text{eff}}$ (Fig.~\ref{sim}a). Therefore, our simulations verifies the theoretical prediction of a statistically significant decrease in anisotropy with increasing $G_{\text{eff}}$, with zippering not affecting this correspondence as theory additionally predicts.
\begin{figure}[t]
\vspace{0cm}
\hspace{-.8cm}
    \centering
    \captionsetup{skip=.2cm,margin={0cm,0cm},justification=raggedright}
    \includegraphics[width=8cm]{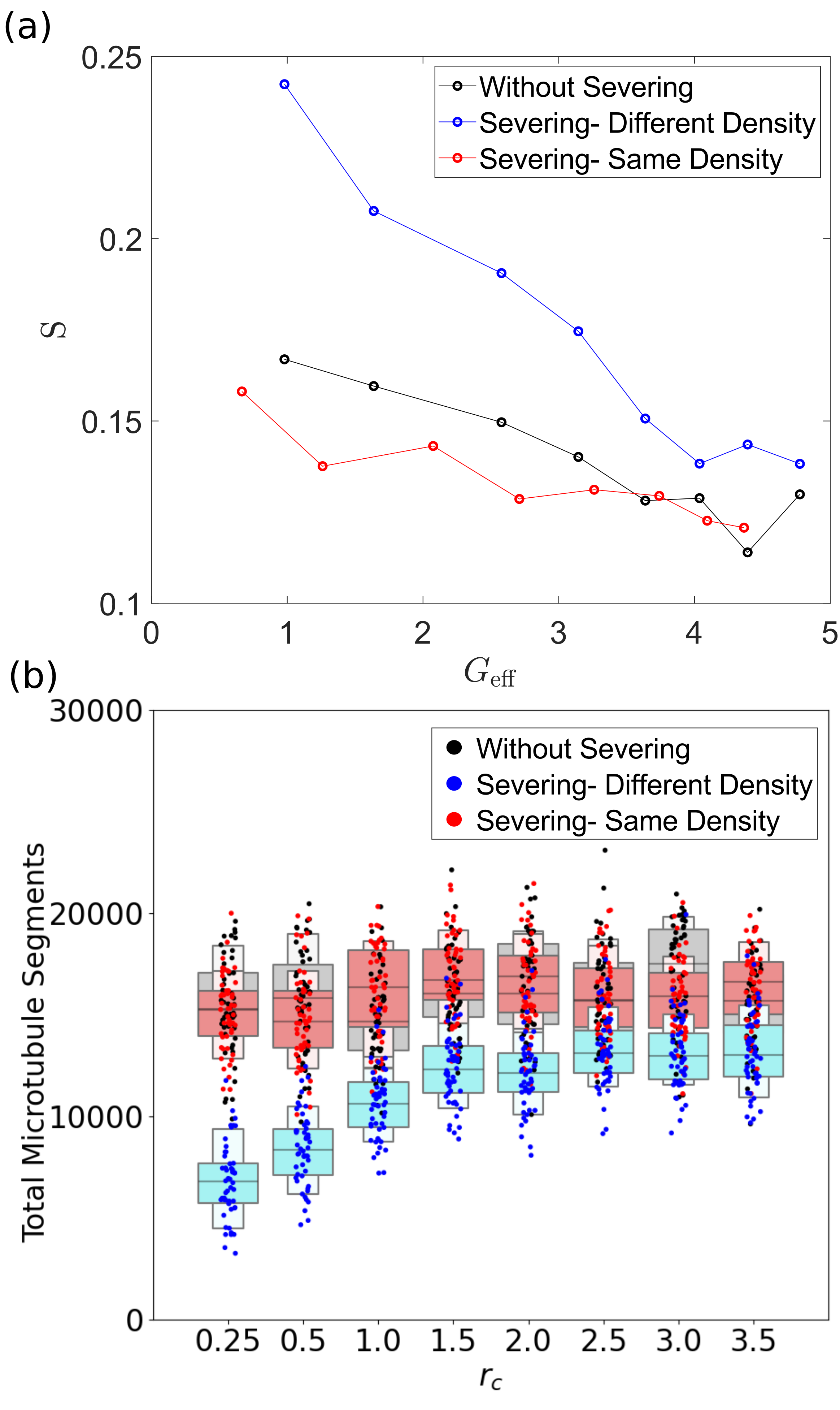}
    \caption{Simulation results showing the effects of severing (zippering is included here). (a) Effect of severing on mean anisotropy as the control parameter $G_\text{eff}$ is increased. (b) Associated number of microtubule segments for each simulation in (a). The case without severing (black) is repeated from Fig.~\ref{fig_6} for comparison. Simulation results with severing included are plotted (blue) as well as simulation results with severing included and the segment density kept constant (red) through changing $r_\text{n}$. Simulation parameter values are provided in Table~\ref{ParamTable}.}
    \label{sev}
\end{figure}
\subsection{Effect of Severing} \label{subsection:Severing}
Through both mean-field theory and simulations, we have shown that crossover severing is not necessary for the anisotropy level of our system to increase as we decrease the control parameter $G_{\text{eff}}$. Introducing the effects of severing into the mean-field theory framework is beyond the scope of this paper. However, it is straightforward to incorporate severing into the simulations to see how it affects the dynamics of the system. For context, we note that that crossover severing has previously been shown experimentally to influence anisotropy magnitudes in microtubule systems \cite{durand,sampathkumar14,lindeboom13}.

Incorporating severing in simulations led to a greater change in anisotropy for the same change in $G_{\text{eff}}$ when compared to the no-severing case, as demonstrated by the steeper gradient in Fig.~\ref{sev}a. However, the inclusion of severing reduced the microtubule density (Fig.~\ref{sev}b) particularly at lower values of $r_{\text{c}}$ which resulted in differing average densities for different values of $r_c$. This coincided with a significant increase in mean anisotropy (Fig.~\ref{sev}a), particularly at lower $G_{\text{eff}}$ values which correspond to lower densities. It has previously been shown in some systems that decreasing microtubule density can increase the order of the system \cite{Chew2023}. As before, we want to keep a similar density across parameter values to ensure a fair comparison. Therefore, we compensate by changing $r_\text{n}$ so that the segment densities remain within a similar range as those of the no-severing simulation (Fig.~\ref{sev}b). After this adjustment, increasing $G_{\text{eff}}$ still causes a decrease in $S$ but at a rate closer to the results from simulations without the adjustment (Fig.~\ref{sev}a).

We conclude that the mean-field theory prediction of anisotropy increasing for decreasing $G_\text{eff}$ is robust to the inclusion of severing. Previous theoretical work in the literature has indicated in 2D that severing can affect microtubule alignment \cite{lou}. Here, the higher increase in anisotropy for decreasing $G_\text{eff}$ when severing is included seems to result from the effect severing has on microtubule segment density. When this effect is adjusted for, we get similar quantitative behavior in the simulation results with or without the inclusion of severing.
\subsection{Comparison to Experiment}
\label{subsection:ExperimentComparison}
Properties of microtubules have been experimentally measured in different systems, with some examples from the literature shown in Table~\ref{TableExprimentalValues}. From these values we can calculate $G_\text{eff}$ using Eq.~\ref{geff}. For example, for tobacco interphase microtubules \cite{tobacco}, we obtain
\begin{equation}
\begin{split}
G_\text{eff}=&  -\left(\frac{4\pi \times 18.36 \upmu \text{m/min} \times 4.59 \upmu \text{m/min}}{18.36\upmu \text{m/min}+4.59\upmu \text{m/min}}\right)^{1/3} \\
& \times \left( \frac{60 \times 0.051/\text{min}}{18.36 \upmu \text{m/min}} -\frac{60 \times 0.015\text{/min}}{4.59 \upmu \text{m/min}}\right) \\
&\times (0.024 \upmu \text{m})^{-1/3} \times (r_{\text{n}})^{-1/3} \\
&=0.37(r_{\text{n}})^{-1/3}. \label{G_effExperiments}
\end{split}
\end{equation}
with the microtubule diameter $d_m$ estimated as $24 \text{nm}$ \cite{arabadopsis}. Here, the negative value of $g$ corresponds to bounded (and therefore physically realistic) growth in the mean-field theory model. Using our prediction of $G^*_\text{eff}=1.56$ from mean-field theory gives an order of magnitude estimate for the nucleation rate at which anisotropy enters the system of $r_{\text{n}} \approx 0.013\text{/}\upmu \text{m}^3\text{/min}$. A similar calculation for another type of tobacco cell \cite{tobacco} and for in vitro experiments \cite{Michison1984} gives $r_n$ estimates of $0.029$ and $0.0073$ respectively (Table~\ref{TableExprimentalValues}). These 3D estimates using experimental data are of comparable order of magnitude to 2D nucleation rates used in simulations of $0.06/\upmu \text{m}^2 / \text{min}$ \cite{comp3} and $0.02 /\upmu \text{m}^2 / \text{min}$ \cite{2} . It is difficult to compare our estimated nucleation rates to nucleation rates observed in experiments from the literature as we have found very few experimental studies which report nucleation rates and when nucleation rates are reported per region they are essentially always in 2D (normally on surfaces). One of the few examples a surface nucleation rate per unit area can be estimated from experiments is Piehl \textit{et. al.} \cite{Piehl2004} where from the measured number of nucleations and centrosome size we estimate $r_\text{n} \approx 40 /\upmu \text{m}^2 / \text{min}$ (although note a centrosome will also give a very different MT structure). This 2D value is significantly above the 3D $r_\text{n}$ threshold for order we calculated for microtubule order in other systems (Table~\ref{TableExprimentalValues}).
\begin{table}[]
    \centering
    \begin{tabular}{c| c  c  c  c | c }
          Source & $v^+$ & $v^-$ & $r_r $ & $r_c $& Prediction \\
         \hline 
         \makecell{Tobacco \\ Interphase \cite{tobacco}} & 4.59 & 18.36 & 0.051 & 0.015 & 0.013 \\ \hline
         \makecell{Tobacco \\ Preprophase \cite{tobacco}} & 6.88 & 17.89  & 0.065  & 0.029  & 0.029  \\ \hline
         \makecell{Suspension \\in vitro \cite{Michison1984}} & 1.9 & 9.7 & $\makecell{\approx0.005 \\\text{estimate}} $ & $\makecell{\approx0.002 \\ \text{estimate}}$ & 0.0073 \\ 
        \hline
    \end{tabular}
\captionsetup{justification=raggedright,belowskip=-.4cm}
     \caption{Experimental values from the literature for microtubule growth conditions and our mean-field theory prediction of $r_\text{n}$ at which the phase transition to order occurs. Growth speeds are stated in units of $\upmu \text{m/min}$, catastrophe and rescue rates in units of /s and $r_\text{n}$ in $/ \upmu\text{m}^{3} \text{/min}$. Estimates of catastrophe and rescue rates from \cite{Michison1984} are obtained by modelling microtubule dynamics as Bernoulli trials, assuming a 0.5 chance of rescue after two shrink events and an average of 10 growth events before a catastrophe to reach $20 \upmu \text{m}$ reported average length.}
    \label{TableExprimentalValues}
\end{table}
\section{Discussion and Outlook}
\label{section:conclusion}
%
In this work, we have developed a 3D mean-field theory model for an interacting system of microtubules. Having established an isotropic solution to this model, we showed that a perturbative solution and therefore a phase transition can only exist for one value of the effective control parameter $G^*_\text{eff}=1.56 $ (3sf). The existence and uniqueness of $G^*_\text{eff}$ was established analytically then its value was numerically calculated with input from experimental estimates for collision event probabilities. We then utilized simulations to verify that anisotropy increased for decreasing $G_\text{eff}$, and the region we observe this decrease coincides with the mean-field theory prediction for the phase transition $G^*_\text{eff}=1.56 $, suggesting this is a reasonable order of magnitude estimate for anisotropy entering the system.

The mean-field theory model furthermore predicts that the critical value $G^*_\text{eff}$ only depends on the induced catastrophe probability function, not the zippering probability function, with our simulations verifying that zippering did not affect how anisotropy depends on $G^*_\text{eff}$. In simulations, crossover severing did not affect the decrease in anisotropy for increasing $G_{\text{eff}}$ provided microtubule density was accounted for. Analyzing the relationship between crossover severing, microtubule density, and anisotropy further, either through extending mean-field theory or experiments, is an interesting future challenge.

There are several ways in which the novel 3D mean-field theory model introduced in this paper could be developed further. Firstly, the effects of crossover severing \cite{severing} could be included within the framework of mean-field theory. This may be possible by altering the length density function by weighting towards shorter microtubule lengths, constrained by the conservation of microtubule length upon splitting. Secondly, a hard boundary could be introduced to the mean-field theory model to allow the study of different cell geometries, although the exact formulation to achieve this is currently unclear. Thirdly, solving the full non-perturbative steady state equations Eqs.~\ref{four} and studying their solution, would lead to an improved understanding of the phase transition taking place. Mean-field theory is an interesting mathematical framework to further explore and analyze microtubules in contrast to computationally intense simulations and time-consuming experiments and is a step to bridge scales from analyzing local microtubule behaviors to multicellular simulations.

\appendix{\textbf{Acknowledgments}}
All three authors acknowledge support from the Gatsby Charitable Foundation (GAT3395/PR4B) and additionally T.A.S and H.J acknowledge support from the Human Frontier Science Program Organization (Grant RGP0009/2018). We thank Fran\c{c}ois N\'{e}d\'{e}lec and B\"{o} Sodeberg for useful discussions in relation to formulating the mean-field theory, as well as Ross Carter for assistance with computational issues.
\\

\appendix{\textbf{Data availability}}
The 3D microtubule simulations were run using Tubulaton
\url{https://gitlab.com/slcu/teamHJ/tubulaton}.

Details of how to reproduce figures in this paper are available here along with the MATLAB scripts
\url{https://gitlab.com/slcu/teamHJ/publications/gibson_etal_2022} .

\appendix

\section{Spherical harmonics and associate Legendre polynomials}
\label{SphericalHarmonics_Legendre}
For completeness, this Appendix provides a full definition of the spherical harmonics and associated Legendre polynomials which are initially introduced in section \ref{lw}. Spherical harmonics \cite{Ferrers,SH} are a complete set of orthogonal functions defined on the surface of a sphere with two indices ($\ell$ and $\text{m}$), which can be defined by the real-valued functions
\begin{equation}
\begin{split}
&Y_{\ell}^\text{m}(\theta,\phi) =\\
&\begin{cases}
 (-1)^\text{m+1}  \sqrt{\frac{2\ell+1}{2 \pi} \frac{(\ell-|\text{m}|)!}{(\ell+|\text{m}|)!}} P_\ell^{-\text{m}} (\cos(\theta)) \sin (\text{m} \phi) &  \text{m} < 0, \\
\sqrt{\frac{2\ell+1}{4 \pi}}P_\ell^\text{0}(\cos(\theta))      &  \text{m} = 0, \\
(-1)^\text{m}  \sqrt{\frac{2\ell+1}{2 \pi} \frac{(\ell-\text{m})!}{(\ell+\text{m})!}} P_\ell^{\text{m}} (\cos(\theta)) \cos (\text{m} \phi)   &  \text{m} >0. \label{sph}
\end{cases}
\end{split}
\end{equation}
Here, $P_\ell^\text{m}$ are the associated Legendre polynomials, defined in terms of the standard Legendre polynomials $P_\ell$ as
\begin{equation}    
    P_{\ell }^{\text{m}}(\cos \theta )=(-1)^\text{m}(\sin \theta )^\text{m} {\frac {d^{\text{m}}}{d(\cos \theta )^{\text{m}}}}\left(P_{\ell }(\cos \theta )\right), \\
\end{equation}
where 
\begin{equation}
  P_\ell(x)=\frac{1}{2^\ell l\text{!}} \frac{d^\ell}{dx^\ell} (x^2-1)^\ell. \label{lpp}  
\end{equation}
\section{Relating isotropic variables to G}
\label{calc3}
Here, proofs of Eq.~\ref{hold} and Eq.~\ref{cubed} are given.

Eq.~\ref{iso4} can be rewritten in the form
\begin{equation}
    \bar{R}=\frac{\bar{L}}{1-4 \pi \alpha \mathfrak{Z}_\text{0} \bar{L} \bar{K}}. \label{root}
\end{equation}
Eq.~\ref{iso3} can also be rewritten as
\begin{equation}
    1+\bar{Q}=\frac{1}{1-4\pi \alpha \mathfrak{Z}_\text{0} \bar{L} \bar{K}}.
\end{equation}
Substituting this expression for $1+\bar{Q}$ into Eq.~\ref{iso2} results in
\begin{equation}
    \bar{K}=\frac{\bar{L}^2}{(1-4 \pi \alpha \mathfrak{Z}_\text{0} \bar{L} \bar{K})^2}. \label{tree}
\end{equation}
Comparing Eq.~\ref{root} and Eq.~\ref{tree} leads to the relation 
\begin{equation}
\bar{K}=\bar{R}^2.
\end{equation}
$\bar{T}$ from Eq.~\ref{iso4} can be substituted into the reciprocal of this identity to obtain
\begin{equation}
    \frac{1}{\bar{K}}=\frac{1}{\bar{R}^2} = \left(\frac{1}{\bar{L}}-4 \pi \alpha \mathfrak{Z}_\text{0} \bar{K} \right)^{2}.
\end{equation}
Substituting for the expression in parentheses using Eq.~\ref{iso1} leads to
\begin{equation} 
\frac{1}{\bar{K}} = (4 \pi \alpha \mathfrak{C}_\text{0} \bar{K} - G)^2  .
\end{equation}
Thus, an expression equivalent to Eq.~\ref{hold} is obtained.
Eq.~\ref{tree} can be rewritten as
\begin{equation}
    \bar{N}(1-4 \pi \alpha \mathfrak{Z}_\text{0} \bar{N})^2=\bar{L}^3. \label{ltee}
\end{equation}
Eq.~\ref{iso1} can also be rewritten
\begin{equation}
    \bar{L}=\frac{(4 \pi \alpha (\mathfrak{Z}_\text{0}+ \mathfrak{C}_\text{0}) \bar{N} -1)^2}{G}.
\end{equation}
Substituting this expression for $\bar{L}$ into Eq.~\ref{ltee} and rearranging for $G^3$ leads to
\begin{equation}
G^3=\frac{(4 \pi \alpha (\mathfrak{Z}_\text{0} +\mathfrak{C}_\text{0}) \bar{N} -1)^3}{\bar{N}(1-4 \pi \alpha \mathfrak{Z}_\text{0} \bar{N})^2}. 
\end{equation}
This is equivalent to Eq.~\ref{cubed}.
\section{Eigenvalue equation for first order perturbations}
\label{calc2}
Here, a proof of Eq.~\ref{GG} is provided. Small perturbations $\lambda(\theta,\phi),\kappa(\theta,\phi),\chi(\theta,\phi), \rho(\theta,\phi)$ to the steady state isotropic system are defined with
\begin{equation}
\begin{split}
&L=\bar{L}(1+\lambda), \\
&K=\bar{K}(1+\kappa),  \\
&Q=\bar{Q}(1+\chi), \\
&R=\bar{R}(1+\rho). \label{firstorder}
\end{split}
\end{equation}
Inserting Eqs. \ref{firstorder} into Eq.~\ref{four2} leads to
\begin{equation}
    \bar{K} +\bar{K} \kappa =(\bar{L}+\bar{L} \lambda)(\bar{R}+\bar{R} \rho)(1+\bar{Q}+\bar{Q} \chi ).
\end{equation}
Substituting for $(1+\bar{Q})$ from Eq.~\ref{iso2} leads to
\begin{equation}
\bar{K} + \bar{K} \kappa = \left(\bar{L}\bar{R}+\bar{L} \bar{R} (\lambda+\rho)+ \mathcal{O}(\lambda \rho) \right) \left(\frac{\bar{K}}{\bar{L} \bar{R}} + \bar{Q} \chi \right).
\end{equation}
Disregarding second (and higher) order terms, this rearranges to
\begin{equation}
    \kappa=\lambda+\rho + \left(\frac{\bar{Q}}{\bar{K}} \bar{L} \bar{R} \chi \right).
\end{equation}
Substituting for $\bar{Q}/\bar{K}$ from Eq.~\ref{iso3} divided by Eq.~\ref{iso2} results in
\begin{equation}
    \kappa=\lambda+\rho + 4 \pi \alpha \mathfrak{Z}_\text{0} \bar{L} \bar{K} \chi. \label{fio4}
\end{equation}
This directly implies that
\begin{equation}
    Z[\kappa]= Z[\lambda+\rho + 4 \pi \alpha \mathfrak{Z}_\text{0} \bar{L} \bar{K} \chi] \label{total1}.
\end{equation}
Inserting Eqs. \ref{firstorder} into \ref{four3} and disregarding second order terms gives
\begin{equation}
    \bar{Q}+\bar{Q} \chi = \bar{L} \bar{K} (1+\bar{Q}) Z[1+\lambda +\kappa +\chi]+\bar{L} \bar{K} \bar{Q} Z[\chi].
\end{equation}
Subtracting Eq.~\ref{iso3} and dividing through by $\bar{Q}$ leaves
\begin{equation}
\chi=\bar{L} \bar{K}\left( \frac{1+\bar{Q}}{\bar{Q}} \right)Z[\lambda+\kappa]+\bar{L} \bar{K}   Z[\chi].
\end{equation}
Substituting for $(1+\bar{Q})/\bar{Q}$ from Eq.~\ref{iso3} gives
\begin{equation}
    \chi=\frac{1}{4 \pi \alpha \mathfrak{Z}_\text{0}} Z[\lambda+\kappa+4 \pi \alpha \mathfrak{Z}_\text{0} \bar{L} \bar{K} \chi]. \label{total2}
\end{equation}
Inserting Eqs. \ref{firstorder} into Eq.~\ref{four4} and subtracting Eq.~\ref{iso4} gives
\begin{equation}
    \rho=\left(\frac{\bar{L}}{\bar{R}}\right)\lambda+ 4 \pi \alpha \mathfrak{Z}_\text{0} \bar{L} \bar{K} (\lambda+\kappa) +\bar{L} \bar{K}Z[\rho].
\end{equation}
Using Eq.~\ref{iso4} again to rewrite the coefficient of $\lambda$ leads to
\begin{equation}
\rho= \lambda + \bar{L} \bar{K} (4 \pi \alpha \mathfrak{Z}_\text{0} \kappa + Z[\rho]). \label{total3}
\end{equation}
From \ref{total1},\ref{total2},\ref{total3} the following expression for $Z[\kappa]$ can be obtained
\begin{equation}
    Z[\kappa]=\frac{1}{2}\left(\frac{\rho-\lambda}{\bar{L} \bar{K}} +4\pi \alpha \mathfrak{Z}_\text{0} (\chi-\kappa)\right). \label{fio1}
\end{equation}
Inserting Eqs. \ref{firstorder} into Eq.~\ref{four1}, subtracting Eq.~\ref{iso1} and taking the first order approximation $(1+\lambda)^{-1} \approx 1-\lambda$ results in
\begin{equation}
   \lambda=-\bar{L}\bar{K}(C[\kappa]+Z[\kappa]). \label{fio2}
\end{equation}
Eq.~\ref{fio1} can be substituted into \ref{fio2} to obtain
\begin{equation}
 2\lambda=-2\bar{L} \bar{K} C[\kappa] - \rho + \lambda - 4 \pi \alpha \mathfrak{Z}_\text{0} \bar{L} \bar{K} \chi + 4\pi \alpha \mathfrak{Z}_\text{0} \bar{L} \bar{K} \kappa.
\end{equation}
Finally, substituting $\kappa$ for the expression in Eq.~\ref{fio4} leads to
\begin{equation}
   -2 \bar{L} \bar{K} C[\kappa] = (1-4 \pi \alpha \mathfrak{Z}_\text{0} \bar{L} \bar{K}) \kappa.
\end{equation}
\section{Variation and convergence of simulations} \label{AppendixC}
Stochastic features the simulations lead to variation between simulations with the same input parameters and fluctuations within any single simulation even at large times when the average behavior appears to have converged. For the with and without zippering simulations, the mean anisotropy shows a clear trend (Fig. \ref{fig_6}) but within each single parameter pair we observe substantial variation in the anisotropy reached, (Fig.~\ref{sim}a). We similarly see variation between simulations in the total number of segments reached, which can in part be attributed to the fluctuations which remain after the average behavior has converged as shown in Fig.~\ref{sim}b. Visually this convergence has occurred well before the $10,000$ time steps limit we use for comparing average behaviors throughout this paper. However, this convergence is slower for lower values of $r_{\text{c}}$ and for $r_{\text{c}}<0.00025$ the average behavior has not converged even by $30,000$ timesteps as indicated by the steadily increasing number of segments (Fig.~\ref{sim}c). As such, we we only analyze simulations with $r_{\text{c}} \ge 0.00025$. Comparing with and without zippering (Fig.~\ref{sim}) we see similar trends as well as no effect on total segment number, further supporting the limited effect of zippering.
\section{Simulation Parameter Values} \label{AppendixD}
Default parameter values used for the Tubulaton simulations are listed here in Table~\ref{ParamTable}.
\renewcommand{\thetable}{A\arabic{table}}
\onecolumngrid
\vspace{.1cm}
\begin{center}
\begin{tabular}{c c} 
 Parameter & Value \\ [0.5ex] 
 \hline\hline
Zippering angle & 0.7 radians   \\ 
Boundary zippering angle threshold &  0.7 radians  \\
Interaction distance  & 49 nm \\
Probability of induced catastrophe  &   $P_\text{c}(\sigma)=\sigma/2 \enspace \forall  \sigma \leq \pi/2 $  and $P_\text{c}(\sigma)=\pi/2 - \sigma/2$ o/w \\
 Probability of Zippering &
 $P_\text{z}(\sigma)=1-P_\text{c}(\sigma) \enspace \forall  \sigma \leq 0.7  \text{, } \sigma \geq \pi-0.7$ and $P_\text{z}(\sigma)=0$ o/w\\
Probability of spontaneous catastrophe & (0.06-3.5) $\times$ $10^{-3}$ per timestep\\
Probability of cutting crossing microtubule  & 0.005\\
Random microtubule shrinkage from either end  & 0 \\
Nucleation rate & (0.145-5.8) $\times$ $10^{-1}$ per timestep \\
Initial nucleations  & 0  \\
 Minus/Plus end shrink/growth speed respectively & 0.08-0.04$\upmu \text{m s}^{-1}$ \\
\hline
 Boundary sphere radius & 1000 units (8 $\upmu$m)\\
 Nucleation sphere radius & 700 units (5.6 $\upmu$m) \\
Analysis sphere radius & 400 units (4 $\upmu$m)  \\
\hline
Number of time steps  & 10,000 \\ 
Number of repeats    & 50 \\
\end{tabular} 
\vspace{-.25cm}
\captionof{table}{List of default parameters for the Tubulaton simulations. \label{ParamTable}}
\end{center}
\renewcommand{\thefigure}{A\arabic{figure}}
\setcounter{figure}{0} 
\begin{figure*}
    \includegraphics[width=17cm]{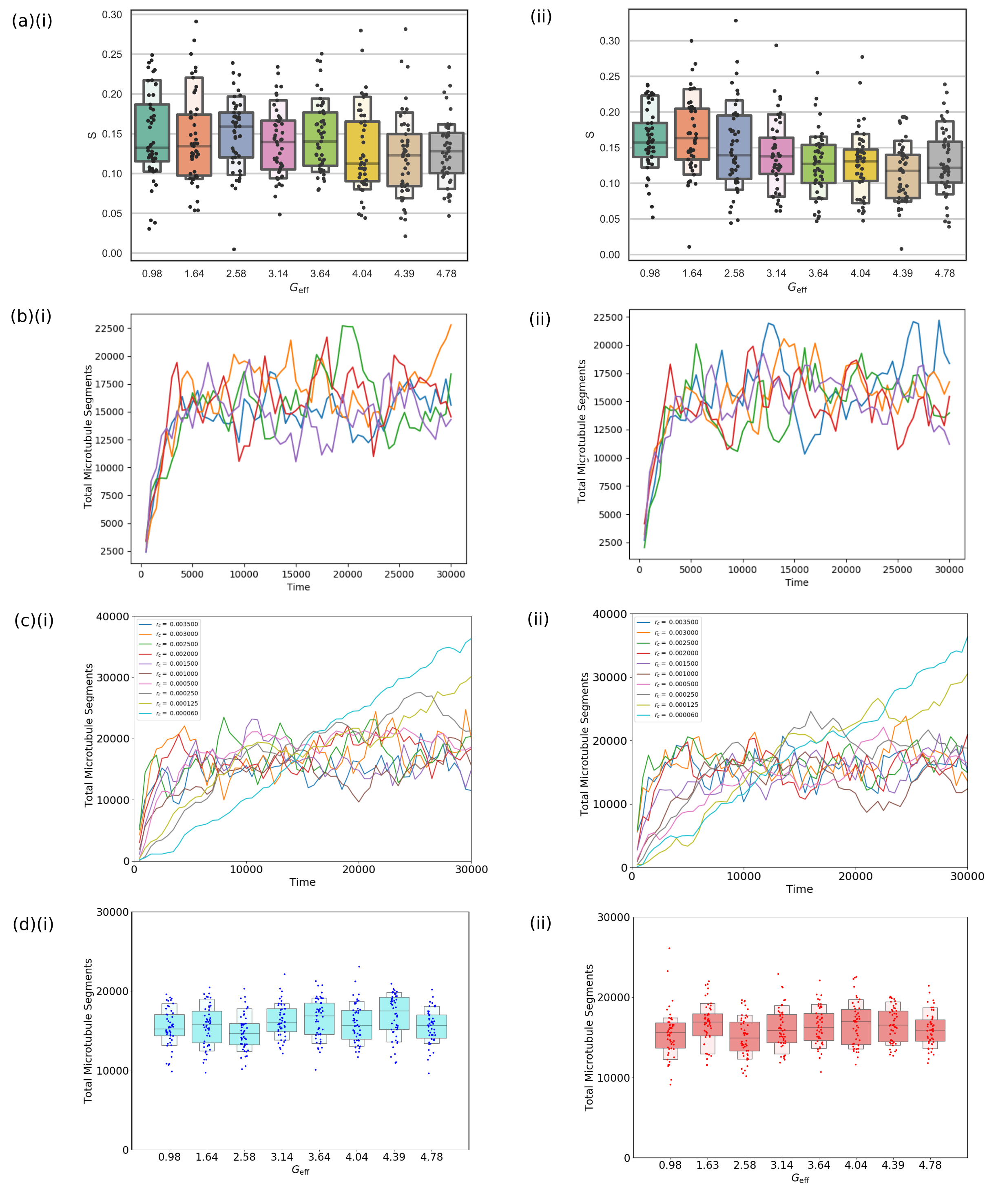}
    \captionsetup{justification=raggedright}
    \caption{Variation in simulation results for (i) with and (ii) without zippering. (a) Boxplots showing variation in anisotropy measure $S$ across all 50 simulation results for each parameter pair (using the data from Fig \ref{fig_6}). (b) Variation in time convergence of total microtubule segments for each of 5 different simulations with same parameters where $r_{\text{c}}=0.0015$. (c) Convergence of total microtubule segments over $30,000$ time steps for different $r_{\text{c}}-r_{\text{n}}$ pairs with $r_{\text{c}}$ in the range from $0.0035$ down to $0.00006$. (d) Total number of microtubule segments in simulations for increasing $G_{\text{eff}}$. Simulation parameter values are provided in Table~\ref{ParamTable}.}
    \label{sim}
\end{figure*}
\clearpage
\twocolumngrid
\bibliography{biblio}

\begin{thebibliography}{10}

\bibitem{micro}
R.~Wade.
\newblock On and around microtubules: An overview.
\newblock {\em Mol. Biotechnol.}, 43:177--91, 2009.

\bibitem{greenbook}
B.~B. Buchanan, W.~Gruissem, and R.~L. Jones.
\newblock {\em Biochemistry and Molecular Biology of Plants}.
\newblock Wiley, 2015.

\bibitem{division}
C.~G. Rasmussen, A.~J. Wright, and S.~Müller.
\newblock The role of the cytoskeleton and associated proteins in determination
  of the plant cell division plane.
\newblock {\em Plant J.}, 75(2):258--269, 2013.

\bibitem{expansion}
L.~Bashline, L.~Lei, S.~Li, and Y.~Gu.
\newblock Cell wall, cytoskeleton, and cell expansion in higher plants.
\newblock {\em Mol. Plant}, 7(4):586--600, 2014.

\bibitem{morpho}
W.~E. Hable, S.~R. Bisgrove, and D.~L. Kropf.
\newblock {To Shape a Plant—The Cytoskeleton in Plant Morphogenesis}.
\newblock {\em Plant Cell}, 10(11):1772--1774, 1998.

\bibitem{5}
A.~Geitmann and A.~Nebenf\"{u}hr.
\newblock Navigating the plant cell: intracellular transport logistics in the
  green kingdom.
\newblock {\em Mol. Biol. Cell}, 26(19):3373--3378, 2015.

\bibitem{6}
M.~Bringmann, B.~Landrein, C.~Schudoma, O.~Hamant, M-T. Hauser, and S.~Persson.
\newblock Cracking the elusive alignment hypothesis: the
  microtubule–cellulose synthase nexus unraveled.
\newblock {\em Trends Plant Sci.}, 17:666 -- 674, 2012.

\bibitem{7}
S.~Reinsch and P.~Gonczy.
\newblock {Mechanisms of nuclear positioning}.
\newblock {\em J. Cell Sci.}, 111(16):2283--2295, 1998.

\bibitem{8}
G.~Schatten, C.~Simerly, and H.~Schatten.
\newblock Microtubule configurations during fertilization, mitosis, and early
  development in the mouse and the requirement for egg microtubule-mediated
  motility during mammalian fertilization.
\newblock {\em Proc. Natl. Acad. Sci. U.S.A.}, 82:4152--6, 1985.

\bibitem{cyto}
R.~H. Goddard, S.~M. Wick, C.~D. Silflow, and D.~P. Snustad.
\newblock {Microtubule Components of the Plant Cell Cytoskeleton}.
\newblock {\em Plant Physiol.}, 104(1):1--6, 1994.

\bibitem{review}
D.~A. Fletcher and R.~D. Mullins.
\newblock Cell mechanics and the cytoskeleton.
\newblock {\em Nature}, 463:485--492, 2010.

\bibitem{dinst}
G.~J. Brouhard.
\newblock Dynamic instability 30 years later: complexities in microtubule
  growth and catastrophe.
\newblock {\em Mol. Biol. Cell}, 26(7):1207--1210, 2015.

\bibitem{angles}
R.~Dixit and R.~Cyr.
\newblock {Encounters between Dynamic Cortical Microtubules Promote Ordering of
  the Cortical Array through Angle-Dependent Modifications of Microtubule
  Behavior[W]}.
\newblock {\em Plant Cell}, 16(12):3274--3284, 2004.

\bibitem{boundary}
C.~J. Ambrose and G.~O. Wasteneys.
\newblock Clasp modulates microtubule-cortex interaction during
  self-organization of acentrosomal microtubules.
\newblock {\em Mol. Biol. Cell}, 19(11):4730--4737, 2008.

\bibitem{severing}
A.~Roll-Mecak and F.~J. McNally.
\newblock Microtubule-severing enzymes.
\newblock {\em Curr. Opin. Cell Biol.}, 22(1):96--103, 2010.
\newblock Cell structure and dynamics.

\bibitem{4}
M.~Piehl, U.~S. Tulu, P.~Wadsworth, and L.~Cassimeris.
\newblock Centrosome maturation: Measurement of microtubule nucleation
  throughout the cell cycle by using gfp-tagged eb1.
\newblock {\em Proc. Natl. Acad. Sci. U.S.A.}, 101(6):1584--1588, 2004.

\bibitem{3}
Y.~Oda.
\newblock Cortical microtubule rearrangements and cell wall patterning.
\newblock {\em Front. Plant Sci.}, 6, 2015.

\bibitem{1}
M.~Yamada and K.~Hayashi.
\newblock Microtubule nucleation in the cytoplasm of developing cortical
  neurons and its regulation by brain-derived neurotrophic factor.
\newblock {\em Cytoskeleton}, 76(5):339--345, 2019.

\bibitem{2}
Y.~Nakaoka, A.~Kimura, T.~Tani, and G.~Goshima.
\newblock {Cytoplasmic Nucleation and Atypical Branching Nucleation Generate
  Endoplasmic Microtubules in Physcomitrella patens  }.
\newblock {\em Plant Cell}, 27(1):228--242, 2015.

\bibitem{branching}
V.~Verma and T.~J. Maresca.
\newblock Direct observation of branching mt nucleation in living animal cells.
\newblock {\em J. Cell Biol.}, 218(9):2829--2840, 07 2019.

\bibitem{liqc}
P.~G.~De Gennes and J.~Prost.
\newblock {\em The Physics of Liquid Crystal}, volume~2.
\newblock Oxford University Press, 1993.

\bibitem{liqcpedagog}
D.~Andrienko.
\newblock Introduction to liquid crystals.
\newblock {\em J. Mol. Liq.}, 267:520--541, 2018.

\bibitem{experiment}
A.~L. Hitt, A.~R. Cross, and R.~C. Williams.
\newblock Microtubule solutions display nematic liquid crystalline structure.
\newblock {\em J. Biol. Chem.}, 265 3:1639--47, 1990.

\bibitem{comp1}
M.~C. Lagomarsino, C.~Tanase, J.~W. Vos, A.~M.~C. Emons, B.~M. Mulder, and
  M.~Dogterom.
\newblock Microtubule organization in three-dimensional confined geometries:
  Evaluating the role of elasticity through a combined in vitro and modeling
  approach.
\newblock {\em Biophys. J.}, 92(3):1046--1057, 2007.

\bibitem{comp2}
E.~C. Eren, R.~Dixit, and N.~Gautam.
\newblock A three-dimensional computer simulation model reveals the mechanisms
  for self-organization of plant cortical microtubules into oblique arrays.
\newblock {\em Mol. Biol. Cell}, 21(15):2674--2684, 2010.

\bibitem{comp4}
V.~A. Baulin, C.~M. Marques, and F.~Thalmann.
\newblock Collision induced spatial organization of microtubules.
\newblock {\em Biophys. Chem.}, 128(2):231--244, 2007.

\bibitem{corticalsimwebsite}
Corticalsim.
\newblock \url{https://github.com/corticalsim/corticalsim}.
\newblock Accessed: September 2022.

\bibitem{comp3}
S.~Tindemans, E.~Deinum, J.~Lindeboom, and B.~M. Mulder.
\newblock Efficient event-driven simulations shed new light on microtubule
  organization in the plant cortical array.
\newblock {\em Front. Phys.}, 2, 2014.

\bibitem{14}
E.~E. Deinum, S.~H. Tindemans, and B.~M. Mulder.
\newblock Taking directions: the role of microtubule-bound nucleation in the
  self-organization of the plant cortical array.
\newblock {\em Phys. Biol.}, 8(5):056002, 2011.

\bibitem{cytosimwebsite}
Cytosim.
\newblock \url{https://gitlab.com/f-nedelec/cytosim}.
\newblock Accessed: September 2022.

\bibitem{11}
B.~Rupp and F.~Nédélec.
\newblock Patterns of molecular motors that guide and sort filaments.
\newblock {\em Lab Chip}, 12:4903--4910, 2012.

\bibitem{tubulatonwebsite}
Tubulaton.
\newblock \url{https://gitlab.com/slcu/teamHJ/tubulaton}.

\bibitem{tubulaton}
V.~Mirabet, P.~Krupinski, O.~Hamant, E.~M. Meyerowitz, H.~Jönsson, and
  A.~Boudaoud.
\newblock {{T}he self-organization of plant microtubules inside the cell volume
  yields their cortical localization, stable alignment, and sensitivity to
  external cues}.
\newblock {\em PLOS Comput. Biol.}, 14:e1006011, 2018.

\bibitem{durand}
P.~Durand-Smet, T.~A. Spelman, E.~M. Meyerowitz, and H.~J{\"o}nsson.
\newblock Cytoskeletal organization in isolated plant cells under geometry
  control.
\newblock {\em Proceedings of the National Academy of Sciences},
  117(29):17399--17408, 2020.

\bibitem{model1}
E.~Geigant, K.~Ladizhansky, and A.~Mogilner.
\newblock An integro-differential model for orientational distributions of
  f-actin in cells.
\newblock {\em SIAM J. Appl. Math}, 59:787--809, 1997.

\bibitem{model2}
I.~S. Aranson and L.~S. Tsimring.
\newblock Theory of self-assembly of microtubules and motors.
\newblock {\em Phys. Rev. E}, 74:031915, 2006.

\bibitem{model3}
V.~Rühle, F.~Ziebert, R.~Peter, and W.~Zimmermann.
\newblock Instabilities in a two-dimensional polar-filament-motor system.
\newblock {\em Eur. Phys. J. E}, 27:243--51, 2008.

\bibitem{model5}
S.~Yarahmadian and M.~Yari.
\newblock Phase transition analysis of the dynamic instability of microtubules.
\newblock {\em Nonlinearity}, 27, 2013.

\bibitem{mamod}
S.~K. Ma.
\newblock {\em Modern theory of critical phenomena}.
\newblock Routledge, 2018.

\bibitem{mft}
M.~Dogterom and S.~Leibler.
\newblock Physical aspects of the growth and regulation of microtubule
  structures.
\newblock {\em Phys. Rev. Lett.}, 70:1347--1350, 1993.

\bibitem{CM}
P.~M. Chaikin and T.~C. Lubensky.
\newblock {\em Mean-field theory}, page 144–212.
\newblock Cambridge University Press, 1995.

\bibitem{Simons}
A.~R. Lamson, J.~M. Moore, F.~Fang, M.~A. Glaser, M.~J. Shelley, and M.~D.
  Betterton.
\newblock Comparison of explicit and mean-field models of cytoskeletal
  filaments with crosslinking motors.
\newblock {\em Eur. Phys. J. E}, 44:1--22, 2021.

\bibitem{mft1}
X.~Q. Shi and Y.~Q. Ma.
\newblock Understanding phase behavior of plant cell cortex microtubule
  organization.
\newblock {\em Proc. Natl. Acad. Sci. U.S.A.}, 107(26):11709--11714, 2010.

\bibitem{baulin}
V.~A. Baulin, C.~M. Marques, and F.~Thalmann.
\newblock Collision induced spatial organization of microtubules.
\newblock {\em Biophys. Chem.}, 128(2):231--244, 2007.

\bibitem{main}
R.~J. Hawkins, S.~H. Tindemans, and B.~M. Mulder.
\newblock Model for the orientational ordering of the plant microtubule
  cortical array.
\newblock {\em Phys. Rev. E}, 82:011911, 2010.

\bibitem{thesis}
P.~Foteinopoulos.
\newblock {\em Models for spatial organization of microtubules and cell
  polarization}.
\newblock PhD thesis, Wageningen University, 2019.

\bibitem{Ferrers}
N.~M. Ferrers.
\newblock {\em An Elementary Treatise on Spherical Harmonics and Subjects
  Connected with Them}.
\newblock Macmillan and Co., London, 1877.

\bibitem{SH}
K.~Atkinson and W.~Han.
\newblock {\em Spherical Harmonics and Approximations on the Unit Sphere: An
  Introduction}.
\newblock Lecture Notes in Mathematics. Springer Berlin Heidelberg, Berlin,
  Heidelberg, 2012 edition, 2012.

\bibitem{LAMY}
X.~Lamy.
\newblock Uniaxial symmetry in nematic liquid crystals.
\newblock {\em Ann. I. H. Poincare-AN}, 32(5):1125--1144, 2015.

\bibitem{orderparam}
H.~L\"owen.
\newblock Anisotropic self-diffusion in colloidal nematic phases.
\newblock {\em Phys. Rev. E}, 59:1989--1995, 1999.

\bibitem{coef}
H.~Wang.
\newblock A new and sharper bound for legendre expansion of differentiable
  functions.
\newblock {\em Appl. Math. Lett .}, 85:95--102, 2018.

\bibitem{antonov}
V.~A. Antonov and K.~V. Holsevnikov.
\newblock An estimate of the remainder in the expansion of the generating
  function for the legendre polynomials (generalization and improvement of
  bernstein’s inequality).
\newblock {\em Vestn. St. Petersbg. Univ.: Math.}, 13:163--166, 1981.

\bibitem{lorch}
L.~Lorch.
\newblock Alternative proof of a sharpened form of bernstein's inequality for
  legendre polynomials.
\newblock {\em Appl. Anal.}, 14(3):237--240, 1983.

\bibitem{ddim}
H.~Kalf.
\newblock {On the expansion of a function in terms of spherical harmonics in
  arbitrary dimensions}.
\newblock {\em Bull. Belg. Math. Soc. Simon Stevin}, 2(4):361 -- 380, 1995.

\bibitem{sampathkumar14}
Arun Sampathkumar, Pawel Krupinski, Raymond Wightman, Pascale Milani, Alexandre
  Berquand, Arezki Boudaoud, Olivier Hamant, Henrik Jönsson, and Elliot~M
  Meyerowitz.
\newblock Subcellular and supracellular mechanical stress prescribes
  cytoskeleton behavior in \textit{Arabidopsis} cotyledon pavement cells.
\newblock {\em eLife}, 3:e01967, 2014.

\bibitem{lindeboom13}
Jelmer~J. Lindeboom, Masayoshi Nakamura, Anneke Hibbel, Kostya Shundyak, Ryan
  Gutierrez, Tijs Ketelaar, Anne Mie~C. Emons, Bela~M. Mulder, Viktor Kirik,
  and David~W. Ehrhardt.
\newblock A mechanism for reorientation of cortical microtubule arrays driven
  by microtubule severing.
\newblock {\em Science}, 342(6163):1245533, 2013.

\bibitem{Chew2023}
Wei-Xiang Chew, Gil Henkin, François Nédélec, and Thomas Surrey.
\newblock Effects of microtubule length and crowding on active microtubule
  network organization.
\newblock {\em iScience}, 26(2):106063, 2023.

\bibitem{lou}
Eva~E. Deinum, Simon~H. Tindemans, Jelmer~J. Lindeboom, and Bela~M. Mulder.
\newblock How selective severing by katanin promotes order in the plant
  cortical microtubule array.
\newblock {\em Proc. Natl. Acad. Sci. U.S.A.}, 114(27):6942--6947, 2017.

\bibitem{tobacco}
P.~Dhonukshe and T.~W.~J. Gadella.
\newblock {Alteration of Microtubule Dynamic Instability during Preprophase
  Band Formation Revealed by Yellow Fluorescent Protein–CLIP170 Microtubule
  Plus-End Labeling[W]}.
\newblock {\em Plant Cell}, 15(3):597--611, 2003.

\bibitem{arabadopsis}
T.~Hashimoto.
\newblock {Microtubules in Plants}.
\newblock {\em The Arabidopsis Book}, 2015(13), 2015.

\bibitem{Michison1984}
T~Mitchison and M.~Kirschner.
\newblock Dynamic instability of microtubule growth.
\newblock {\em Nature}, 312:237--242, 1984.

\bibitem{Piehl2004}
M.~Piehl, U.~S. Tulu, P.~Wadsworth, and L.~Cassimeris.
\newblock Centrosome maturation: Measurement of microtubule nucleation
  throughout the cell cycle by using gfp-tagged eb1.
\newblock {\em Proc. Natl. Acad. Sci. U.S.A.}, 101(6):1584--1588, 2004.

\end{thebibliography}
\bibliographystyle{unsrt}
\end{document}